\documentclass{aa}  

\RequirePackage{etex}

\usepackage{graphicx,comment}
\usepackage{txfonts}
\usepackage{newtxtext,newtxmath}
\usepackage{amsmath}	
\usepackage[percent]{overpic}

\usepackage{anyfontsize}
\usepackage{xspace}
\usepackage{amssymb}
\usepackage{amsmath} 
\usepackage{graphicx}
\usepackage{txfonts}
\usepackage{lipsum}
\usepackage{longtable}
\usepackage{natbib}
\usepackage{comment}

\usepackage{hyperref}
\hypersetup{
    colorlinks=true,
    linkcolor=blue,
    citecolor=blue,
    urlcolor=blue
}

\newcommand{\thisgrb}{GRB 220107A\xspace}
\newcommand{\astrosat}{{\em AstroSat}\xspace}
\newcommand{\fermi}{{\em Fermi}\xspace}
\newcommand{\kw}{{\em Konus}-{Wind}\xspace}

\newcommand{\fermiT}{{T$_{\rm 0}$}\xspace}

\newcommand{\swift}{{\em Swift}\xspace}
\newcommand{\tninty}{{$T_{\rm 90}$}\xspace}

\newcommand{\Ep}{$E_{\rm p}$\xspace}

\begin{document} 

\title{Episode-wise spectro-polarimetry of GRB 220107A: Testing the hypothesis of evolving radiation mechanisms}

\author{Rahul Gupta\inst{1}\fnmsep\thanks{Corresponding author: rahulbhu.c157@gmail.com, rahul.gupta@nasa.gov}\fnmsep\thanks{NASA Postdoctoral Program Fellow}
\and Rushikesh Sonawane \inst{2, 3}
\and Shabnam Iyyani \inst{2,3}
\and D. Frederiks \inst{4}
\and Judith Racusin \inst{1}
\and Tanmoy Chattopadhayay \inst{5}
\and A. J. Castro-Tirado \inst{6, 7}
\and A. F. Valeev \inst{8, 9}
\and Soumya Gupta \inst{10, 11}
\and Mayuresh Tembhurnikar \inst{12}
\and A. Ridnaia \inst{4}
\and D. Svinkin \inst{4}
\and S. B. Pandey \inst{13}
\and Dipankar Bhattacharya\inst{14}
\and Vidushi Sharma \inst{1, 15}
\and Varun Bhalerao\inst{16}
\and G. C. Dewangan \inst{12}
\and Santosh Vadawale\inst{17}
\and R. S{\'a}nchez-Ram{\'i}rez \inst{6}
\and Anastasia Tsvetkova\inst{4, 18}
}

\institute{Astrophysics Science Division, NASA Goddard Space Flight Center, Mail Code 661, Greenbelt, MD 20771, USA
\and Centre for High Performance Computing, Indian Institute of Science Education and Research Thiruvananthapuram, Thiruvananthapuram, 695551, India
\and School of Physics, Indian Institute of Science Education and Research Thiruvananthapuram, Thiruvananthapuram, 695551, India
\and Ioffe Institute, 26 Politekhnicheskaya, St. Petersburg, 194021, Russia
\and Kavli Institute of Particle Astrophysics and Cosmology, Stanford University, 452 Lomita Mall, Stanford, CA 94305, USA 
\and Instituto de Astrof\'isica de Andaluc\'ia (IAA-CSIC), Glorieta de la Astronom\'ia s/n, E-18008, Granada, Spain
\and Ingenier\'ia de Sistemas y Autom\'atica, Universidad de M\'alaga, Unidad Asociada al CSIC por
el IAA, Escuela de Ingenier\'ias Industriales, Arquitecto Francisco
Pe\~nalosa, 6, Campanillas, 29071 M\'alaga, Spain
\and Special Astrophysical Observatory of Russian Academy of Sciences, Nizhniy Arkhyz, Russia 
\and Crimean Astrophysical Observatory, Russian Academy of Sciences, Nauchnyi, 298409 Russia  
\and Bhabha Atomic Research Center, Mumbai, Maharashtra-400094, India
\and Homi Bhabha National Institute, Mumbai, Maharashtra-400094, India 
\and Inter-University Center for Astronomy and Astrophysics, Pune, Maharashtra 411007, India
\and Aryabhatta Research Institute of Observational Sciences (ARIES), Manora Peak, Nainital-263002, India
\and Department of Physics, Ashoka University, Sonipat, Haryana-131029, India  
\and Center for Space Science and Technology, University of Maryland Baltimore County, Baltimore, MD 21250, USA
\and Indian Institute of Technology Bombay Powai, Mumbai, Maharashtra 400076, India 
\and Physical Research Laboratory, Ahmedabad, Gujarat 380009, India
\and Dipartimento di Fisica, Università degli Studi di Cagliari, SP Monserrato-Sestu, km 0.7, I-09042 Monserrato, Italy
}



\abstract
    {The radiation mechanisms powering the prompt emission of GRBs remain a long-standing question, with both synchrotron and photospheric models offering plausible explanations. Time-resolved spectro-polarimetric measurements provide powerful diagnostics of the emission processes.} 
    {We investigate the spectro-polarimetric properties of the long-duration GRB~220107A, which exhibited two distinct emission episodes separated by a $\sim$40 s quiescent gap, to test whether such multi-episode bursts show evidence for evolution in their underlying radiation mechanisms.}
   {We analyzed prompt emission data from \astrosat/CZTI, \fermi/GBM, and \kw, performing spectro-polarimetric analysis for each emission episode. In addition, we report a redshift of $z$ = 1.246 using 6\,m BTA telescope.}   
   {The time-integrated polarization analysis (\fermiT-2 to \fermiT+106~s) shows no significant detection (${\rm PF} < 38\%$, $2\sigma$ confidence; ${\rm BF} = 0.64$). Time-resolved analysis reveals clear spectral evolution between the two episodes, with episode 1 exhibiting a hard low-energy photon index and episode 2 showing substantial spectral softening ($\alpha \sim -0.72$). Regarding polarization: Episode 1 shows a low polarization upper limit ($1.5\sigma$ upper limit $< 52\%$), consistent with expectations for photospheric emission dominated by quasi-thermal Comptonization in a baryon-rich outflow. Episode 2 also shows overall low polarization (${\rm PF} < 55\%$, $2\sigma$; ${\rm BF} \sim 1$), though sliding-window analysis yields a marginally elevated signal (${\rm PF} = 70 \pm 30\%$, ${\rm BF} = 2.8$) between \fermiT+76 to \fermiT+88 s. We interpret these measurements cautiously: The robust spectral softening between episodes could arise from sub-photospheric dissipation, optically thin synchrotron radiation in small-scale magnetic fields, or if the tentative polarization enhancement proves intrinsic, it would favor synchrotron emission in large-scale ordered magnetic fields.}      
   {The spectral evolution of GRB 220107A, combined with our polarimetric constraints, demonstrates the diagnostic potential of time-resolved spectro-polarimetry for constraining GRB prompt emission physics. While our polarization measurements remain below definitive detection thresholds, we present GRB 220107A as a test case illustrating how future higher sensitivity observations could discriminate between competing emission models for multi-episode bursts. Our results emphasize both the promise and current limitations of prompt phase polarimetry.}

\titlerunning{Spectro-polarimetric evolution of GRB 220107A}

\maketitle

\section{Introduction}

Gamma-ray bursts (GRBs) represent the most energetic electromagnetic transients in the Universe, releasing 10$^{51}$-10$^{54}$ erg within seconds to minutes \citep{2004IJMPA..19.2385Z, 2004RvMP...76.1143P}. Despite decades of observations, the dominant radiation mechanism responsible for their prompt emission remains one of the most enduring puzzles in high-energy astrophysics \citep{2022Galax..10...38B, 2011CRPhy..12..206Z}. The observed prompt emission spectra, typically characterized by a smoothly broken power-law extending from keV to MeV energies, have been interpreted through two competing theoretical frameworks: synchrotron emission \citep{2020NatAs...4..174B, 2019A&A...625A..60R, 2018MNRAS.476.1785B}, and photospheric emission \citep{1986ApJ...308L..47G, 2017SSRv..207...87B, 2015ApJ...802..134B, 2020ApJ...893..128A}.

The synchrotron shock model, originally proposed as the standard framework for GRB prompt emission \citep{Rees&Meszaros1992, 1994ApJ...430L..93R}, invokes internal collisions between shells with different Lorentz factors ($\Gamma$ $\sim$ 100-1000) within the relativistic jet. In this scenario, electrons are accelerated to non-thermal distributions via Fermi acceleration in collisionless shocks, subsequently radiating synchrotron emission in the post-shock magnetic field \citep{1994ApJ...430L..93R, 2009A&A...498..677B, 2020NatAs...4..210Z}. However, this model faces significant challenges: the observed low-energy spectral indices often violate the synchrotron ``line of death" ($\alpha$ $>$ -2/3; \citealt{1997ApJ...479L..39C, 1998ApJ...506L..23P, 2002ApJ...581.1248P}), and the typical peak energies require extreme magnetic field amplification or electron acceleration efficiencies \citep{1994ApJ...432L.107K, 2021ApJ...913...60P, 2007Ap&SS.309..157D}. Recent theoretical developments have explored synchrotron emission in decaying magnetic fields, marginally fast cooling regimes, and magnetic reconnection sites as potential solutions to these discrepancies \citep{2014NatPh..10..351U, 2014ApJ...780...12Z, 2020NatAs...4..174B}.

The photospheric emission model offers an alternative paradigm where the observed radiation originates from the photosphere of the ultra-relativistic fireball—the surface where optical depth $\tau$ $\sim$ 1 \citep{2015AdAst2015E..22P}. In the pure photospheric scenario, the emission emerges as a quasi-thermal spectrum, potentially modified by geometric effects, viewing angle dependencies, and velocity gradients within the outflow \citep{2013A&A...551A.124H, 2011ApJ...732...34L, 2011ApJ...732...49P, 2018ApJ...853....8P}. However, pure thermal emission struggles to explain the broad spectral peaks and non-thermal power-law extensions observed in many bursts. This has motivated hybrid models incorporating sub-photospheric dissipation, where energy release below the photosphere (via shocks, magnetic reconnection, or neutron-proton collisions) modifies the emergent spectrum through inverse Compton scattering and photon breeding, producing a non-thermal component superimposed on the thermal seed \citep{2005ApJ...628..847R, 2006ApJ...642..995P, 2006A&A...457..763G, 2010MNRAS.407.1033B}. These models have been further developed to include geometric effects and viewing angle dependencies, which are critical for interpreting polarimetric data \citep{2011ApJ...732...49P, 2013A&A...551A.124H}.

While spectroscopy has traditionally served as the primary diagnostic tool for GRB emission mechanisms \citep{2004ApJ...613..460B, 2023MNRAS.519.3201C, 2024A&A...683A..55C, 2021MNRAS.505.4086G}, spectral analysis alone suffers from fundamental degeneracies. Both synchrotron and modified photospheric models can produce statistically acceptable fits to the same observed spectra \citep{2018JApA...39...75I}. The Band function \citep{Band1993}, empirically describing most GRB spectra, can be interpreted as synchrotron emission with specific electron distributions or as Comptonized photospheric emission with particular dissipation profiles.

Prompt emission polarimetry provides a crucial independent constraint that can break these degeneracies \citep{2021Galax...9...82G}. The polarization fraction (PF) and polarization angle (PA) are sensitive to the underlying physics of the emitting region, including the geometry of the outflow, the structure of the magnetic field, and the nature of the emission process \citep{2009ApJ...698.1042T, 2014MNRAS.440.3292L, 2017NewAR..76....1M}. Synchrotron emission from ordered magnetic fields naturally produces high linear polarization, with theoretical predictions reaching $\sim$ 60-70$\%$ for uniform fields perpendicular to the line of sight. The observed polarization depends critically on the magnetic field configuration: globally ordered fields (e.g., large-scale toroidal fields generated by jet rotation) yield high net polarization, while tangled fields on scales smaller than the emission region result in depolarization through averaging \citep{2021MNRAS.504.1939G}. In contrast, photospheric emission inherently produces low polarization (typically $\leq$ 10$\%$) due to multiple scatterings in the optically thick regime, though geometric effects can introduce modest polarization \citep{2014MNRAS.440.3292L, 2018ApJ...856..145L}.

The detection of gamma-ray polarization has been historically challenging due to the transient nature of GRBs and the technical difficulties of hard X-ray/gamma-ray polarimetry \citep{2021JApA...42..106C, 2022hxga.book...33B}. Early measurements, including observations of GRB~021206 by the Reuven Ramaty High Energy Solar Spectroscopic Imager (RHESSI; 80$\pm$20\%, \citealt{2003Natur.423..415C}), suggested high polarization but suffered from large systematic uncertainties \citep{2004MNRAS.350.1288R}. The Imager on Board the INTEGRAL Satellite / INTEGRAL Soft Gamma-Ray Imager (IBIS/ISGRI) and the Spectrometer on INTEGRAL (SPI) instruments aboard the INTErnational Gamma-Ray Astrophysics Laboratory (INTEGRAL) provided constraints for several bursts \citep{2021arXiv211207644G}, while dedicated polarimeters such as the Gamma-Ray Burst Polarimeter (GAP) aboard Interplanetary Kite-craft Accelerated by Radiation Of the Sun (IKAROS; \citealt{2011PASJ...63..625Y, 2011ApJ...743L..30Y}) and the POLAR instrument \citep{2020A&A...644A.124K} have expanded the sample, though with mixed results ranging from null detections to a high degree of time-integrated polarization.

Significant progress in GRB polarization studies has been enabled by the Cadmium Zinc Telluride Imager (CZTI) aboard \astrosat, which operates as a hard X-ray polarimeter in the 100-600 keV range through Compton scattering and measures \emph{linear} polarization \citep{2022ApJ...936...12C}. In this work, we consider only linear polarization; circular polarization is not discussed, and the term ``polarization" hereafter refers exclusively to linear polarization. \astrosat/CZTI has now measured time-integrated polarization for over two dozen GRBs \citep{2022ApJ...936...12C}. However, most GRB light curves consist of overlapping pulses, making it difficult to isolate distinct emission regimes. Time-integrated polarization analyses yield global constraints on the jet properties, while time-resolved measurements reveal potential evolution of emission regions, radiation processes, and magnetic field structures during the burst, offering deeper insights into the emission mechanisms. Therefore, a detailed time-resolved spectro-polarimetric analysis is attempted for bright \astrosat GRBs to confirm their radiation mechanisms \citep{2019ApJ...874...70C, 2018ApJ...862..154C, 2019ApJ...882L..10S, 2022MNRAS.511.1694G, 2024ApJ...972..166G, 2025A&A...701A.172G, 2025ApJ...990..215S, 2020MNRAS.493.5218S}. Particularly intriguing are hints of time-variable polarization within individual bursts, suggesting evolving magnetic field configurations or changing emission mechanisms during the prompt phase \citep{2011ApJ...743L..30Y, 2005A&A...439..245W, 2019ApJ...882L..10S, 2020MNRAS.493.5218S, 2024ApJ...972..166G, 2025A&A...701A.172G, 2025ApJ...990..215S, McGlynn2007...466..895M, gotz2009_INTEGRAL_pol}.

In this paper, we present an episode-wise spectro-polarimetric analysis (time-integrated and time-resolved) of GRB 220107A. This burst exhibits two distinct emission episodes separated by a $\sim 40$ s quiescent gap, offering an ideal laboratory to test for the evolution of radiation mechanisms within a single event and evaluate the diagnostic power of spectro-polarimetry. By combining time-resolved hard X-ray polarimetry from \astrosat/CZTI with broadband spectroscopy from \kw and \fermi/GBM, we test the hypothesis that the dominant emission mechanism transitions from a photospheric process in the first episode to synchrotron emission in the second. Section ~\ref{sec:grb220107a} provides an overview of GRB 220107A's observational properties and multi-wavelength detections. Section ~\ref{sec:data} describes the data reduction and analysis methods for \astrosat-CZTI, \fermi-GBM, and \kw observations. Section ~\ref{sec:results} presents our results, including time-integrated and episode-wise spectral fits, polarization measurements and their temporal evolution, as well as the redshift measurement and its prompt correlation. Section ~\ref{sec:discussion} discusses the implications for emission mechanisms and jet compositions. We examine how the evolving polarization signatures discriminate between competing models. Finally, Section ~\ref{sec:conclusion} summarizes our conclusions and their broader implications for understanding GRB radiation mechanisms.

\section{GRB~220107A observations}
\label{sec:grb220107a}

\subsection{Prompt emission observations}
\begin{figure}[!ht]
\centering
\includegraphics[width=\hsize]{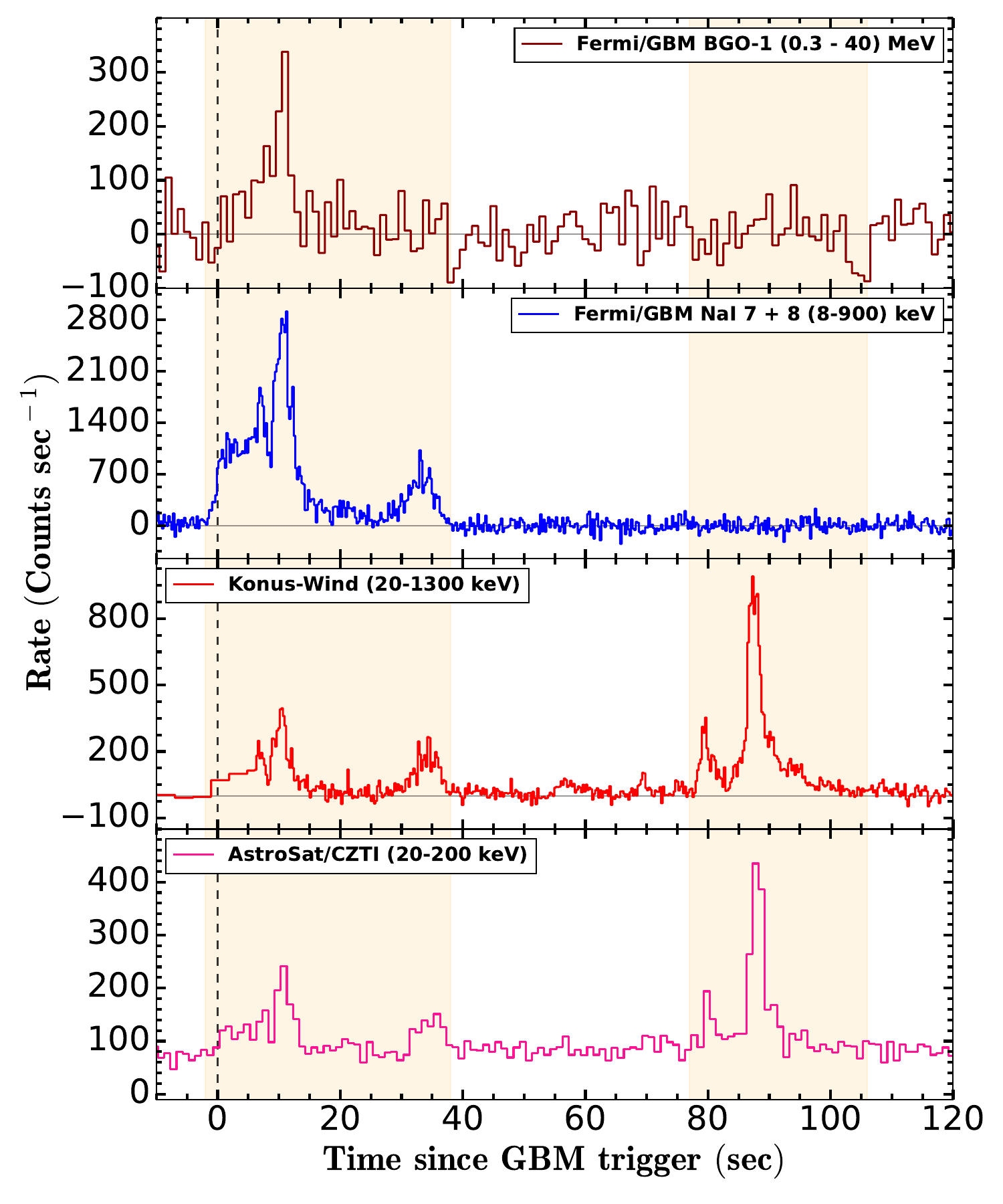}
\caption{Prompt light curve of GRB~220107A from multiple observatories across various energy bands. Both \astrosat/CZTI and \kw detected emission from the first and second episodes of the burst, whereas \fermi/GBM detected only the initial episode due to Earth occultation. The \kw light curve aligns with \astrosat/CZTI observations, confirming emission during the period when GBM was obscured. The vertical line marks the GBM trigger time, and the shaded regions indicate the durations of the first (\fermiT-2 to \fermiT+ 38 sec) and second (\fermiT+77 to \fermiT+ 106 sec) episodes.}
\label{fig:promptlc}
\end{figure}

GRB 220107A was detected by the Gamma-ray Burst Monitor (GBM) aboard the \fermi satellite \citep{2009ApJ...702..791M} at 14:45:31.93 UT on 2022 January 7 (\fermiT; \citealt{2022GCN.31399....1F}). 
The GBM temporal profile exhibits several overlapping episodes, yielding a \tninty of approximately 33 s in the 50--300 keV band (see top two panels of Figure \ref{fig:promptlc}). 
Additional prompt detections occurred with other high-energy observatories, such as \astrosat/CZTI \citep{2022GCN.31408....1S}, \kw \citep{2022GCN.31429....1R} and \swift/BAT-GUANO \citep{2022GCN.31402....1D}. \astrosat/CZTI shows two emission episodes (a weaker episode followed by a brighter episode) and measured \tninty of 97$^{+4}_{-5}$ s in the 20--200 keV band \citep{2022GCN.31408....1S}. \kw light curve also shows two emission episodes consistent with \astrosat/CZTI (see Figure \ref{fig:promptlc}) and provides complementary spectral coverage. 

\begin{figure}[!ht]
\centering
\includegraphics[scale=0.39]{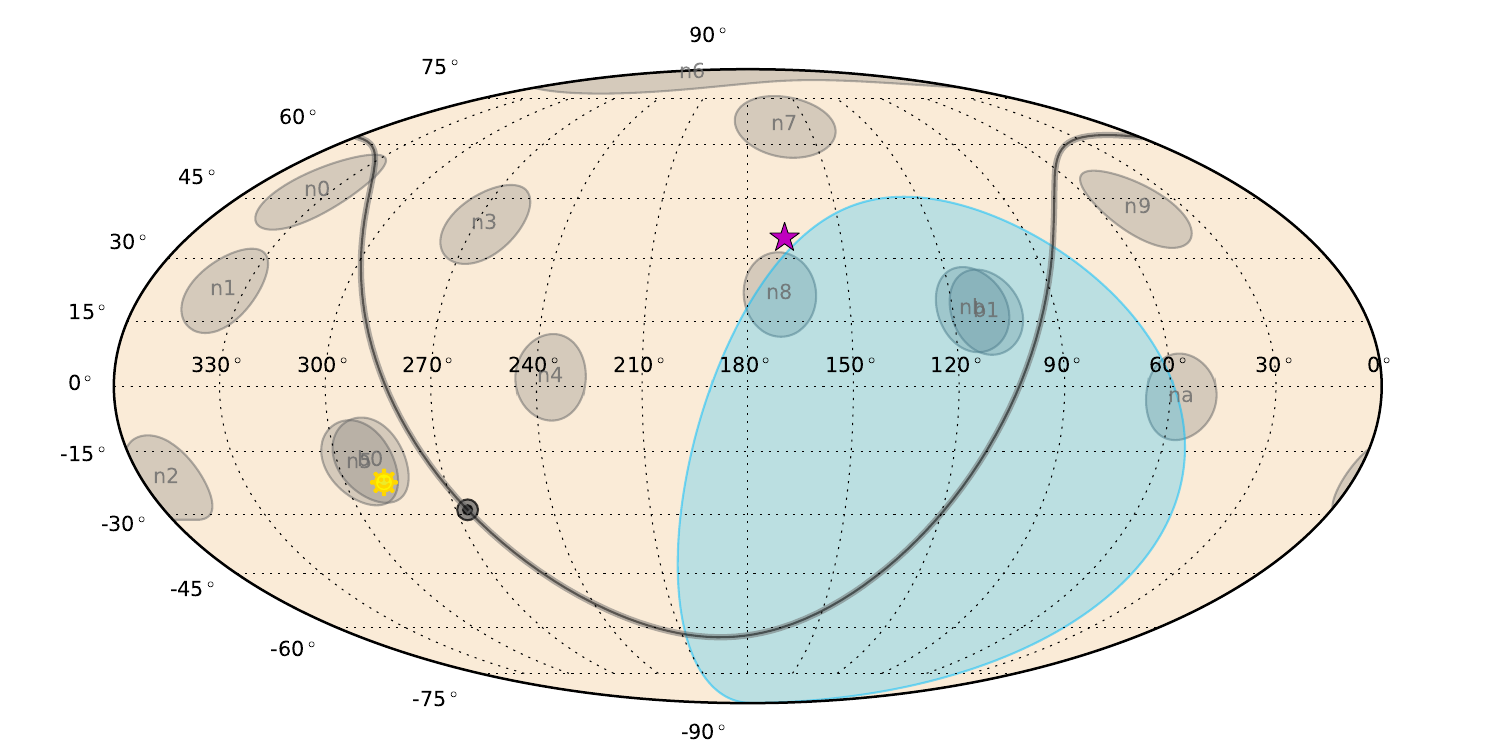}
\includegraphics[scale=0.35]{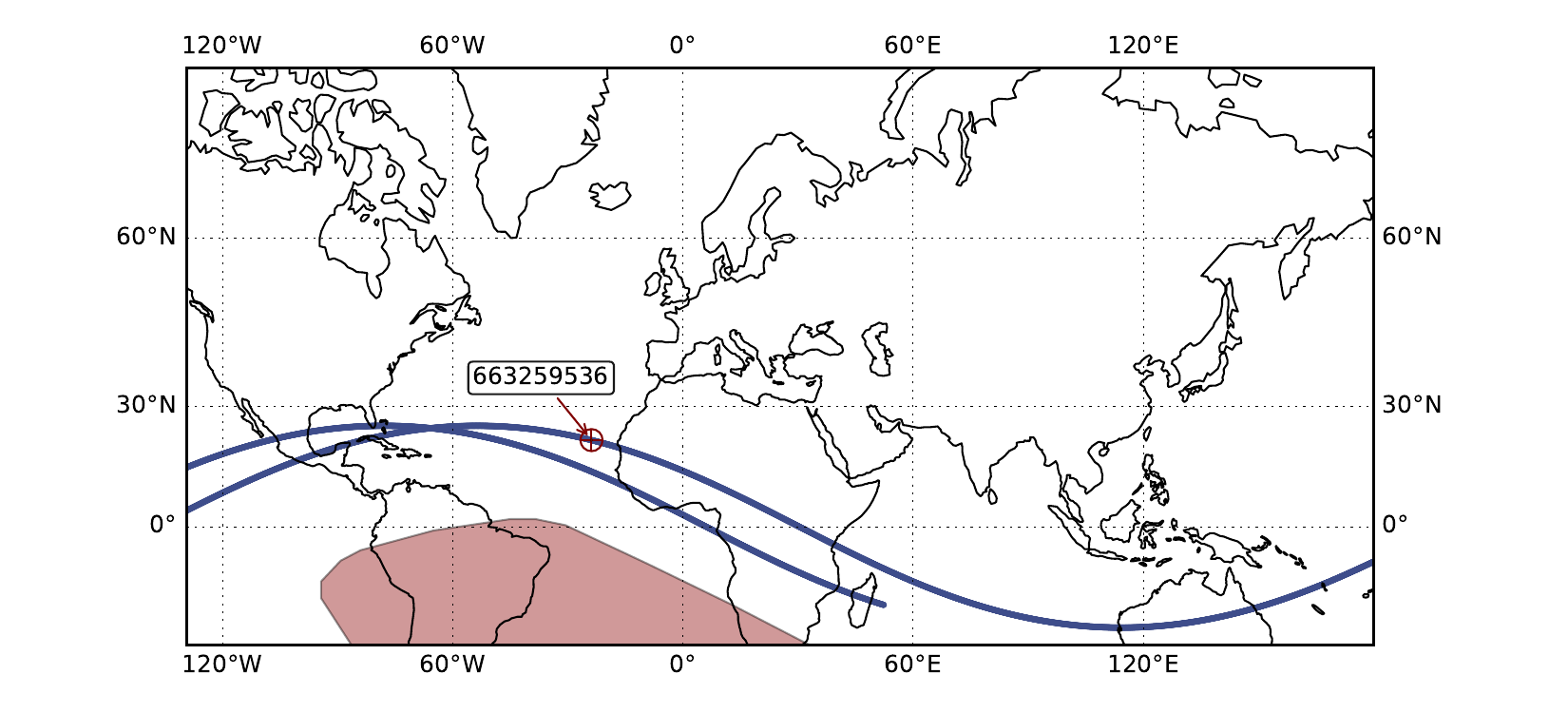}
\caption{Analysis of the \fermi/GBM field of view for GRB 220107A, illustrating the visibility of the GRB position relative to Earth occultation. The upper panel shows a full-sky map in celestial coordinates showing the location of GRB 220107A, marked by a pink star. The pointing directions of all 14 GBM detectors at the trigger time are indicated by light-gray circles, which represent the detector normal directions and do not correspond to the detectors' fields of view. The Galactic plane is shown as a thick gray band, with the Galactic center marked by a circle. The Earth, as seen from the \fermi spacecraft in orbit, is shown as a blue polygon, indicating the Earth-occulted region of the sky. The position of the Sun is marked by a yellow smiley symbol. The lower panel shows the geolocation of the \fermi spacecraft at the time of the trigger, marked by a red circle with the mission elapsed time (MET). The red shaded region highlights the South Atlantic Anomaly (SAA), while the solid blue line indicates the orbital trajectory of the \fermi spacecraft.}
\label{fig:GBMFOV}
\end{figure}

Our analysis of the \fermi/GBM light curve for GRB 220107A indicates a weaker temporal episode initially, with a notable absence of detectable emission during the main and most intense episode of the burst. We investigated the visibility of the GRB's sky position relative to the \fermi/GBM detectors at the time of the event. As illustrated in Figure~\ref{fig:GBMFOV}, the GRB's position was not obscured by Earth during the onset of the first episode. However, during the second episode, the GRB's coordinates were outside the field of view of the GBM detectors due to Earth occultation. This accounts for the absence of detectable emission in the GBM light curve during the second episode of GRB 220107A, as corroborated by simultaneous observations from the \kw and \astrosat/CZTI, which detected emission during this phase. At the trigger time, the closest GBM NaI detector was NaI 8, positioned at an angle of 13.7$^\circ$ from the GRB's location.

\subsection{Soft X-ray observations}
We analyzed the X-ray afterglow of GRB~220107A using observations from the \swift X-Ray Telescope (XRT; \citealt{2005SSRv..120..165B}), which was triggered (at $T_0+26.7$ ks) through a Target of Opportunity (ToO) program to follow up the \swift/BAT-GUANO-detected burst \citep{2022GCN.31402....1D}. An uncatalogued X-ray source identified as the X-ray afterglow of GRB~220107A at RA (J2000) = 11h 19m 13.55s, Dec (J2000) = +34d 10' 12.5$^"$ with a 90\% error radius of 2.3$^"$ \citep{2022GCN.31410....1B}.

\subsection{\swift/UVOT and Optical observations}
The \swift Ultraviolet and Optical Telescope (UVOT; \citealt{2005SSRv..120...95R}) initiated settled observations of the field of GRB 220107A's field 26.7 ks post-\fermi trigger \citep{2022GCN.31415....1S}, detecting an uncatalogued fading source aligned with the GUANO localization and XRT position \citep{2022GCN.31410....1B}. 
Initial UVOT white filter magnitudes from 26.7--107 ks post-trigger range from 20.25 $\pm$ 0.08 to 21.57 $\pm$ 0.20 (Vega system, uncorrected for Galactic extinction of $E(B-V) = 0.056$ \citep{1998ApJ...500..525S}). Ground-based follow-up included MITSuME Akeno observations starting 5.4 hours post-trigger, detecting the source at RA (J2000) = 11h 19m 13.86s, Dec (J2000) = +34d 10' 14.85$^"$ (offset 4.4 arcsec from the XRT position) with magnitudes $g' = 19.8 \pm 0.5$, $R_c = 18.9 \pm 0.2$, and $I_c = 18.3 \pm 0.2$ \citep{2022GCN.31413....1H}. Nanshan/NEXT imaging at 29.46 hours yielded $r = 20.22 \pm 0.05$ \citep{2022GCN.31416....1Z}. Further detections reported from SAO RAS 1-m Zeiss-1000 on January 9 ($T_{\rm mid}$-\fermiT = 1.496--1.529 days) with $R_c = 20.41 \pm 0.03$, $V = 20.97 \pm 0.06$, and $B = 22.08 \pm 0.18$ \citep{2022GCN.31422....1M}; AbAO AS-32 and Terskol K-800 on January 8--9 ($T_{\rm mid}-T_0 = 1.386$--$1.486$ days) with $R = 20.26 \pm 0.14$ and clear $= 20.24 \pm 0.14$ \citep{2022GCN.31419....1P}; additional SAO RAS $R_c$-band data on January 9 ($T_{\rm mid}-T_0 = 2.336$ days) at $R_c = 21.01 \pm 0.04$ \citep{2022GCN.31426....1M}; and SAO RAS $R_c$-band data on January 10--11 ($T_{\rm mid}-T_0 = 3.416$--$3.517$ days) at $R_c = 21.69 \pm 0.10$ and $22.03 \pm 0.09$ \citep{2022GCN.31431....1M}. We utilized available UVOT and optical photometric measurements from these GCN reports to derive the temporal evolution of the optical afterglow.

\section{Data Analysis}
\label{sec:data}

\subsection{Temporal and Spectral data analysis}
\label{spec_temp}

The prompt emission characteristics of GRB~220107A were extracted using data from the \fermi/GBM, \astrosat/CZTI, and \kw instruments. For temporal analysis of \fermi/GBM data, background-subtracted light curves were generated using the GBM data tools\footnote{\url{https://fermi.gsfc.nasa.gov/ssc/data/analysis/gbm/gbm_data_tools/gdt-docs/}} in different energy bands. The sodium iodide (NaI 7 and NaI 8) and bismuth germanate (BGO 1) detectors were selected for this analysis due to their favorable viewing angles relative to the burst, which optimized the signal-to-noise ratio. The resulting light curves are presented in Figure~\ref{fig:promptlc}: panel~2 (blue) illustrates the combined NaI~7 and NaI~8 light curve in the 8--900~keV energy range, while panel~1 (dark red) shows the BGO~1 light curve in the 0.3--40~MeV range. 
\kw observations (panel~3) in the 20--1300~keV energy band clearly resolve two distinct emission episodes. The last panel of Figure~\ref{fig:promptlc} shows the \astrosat/CZTI single-event light curve in the 20--200~keV range. Orange shaded regions mark the time interval used for the episode-wise polarization analysis.

The spectral properties of GRB~220107A were investigated using prompt emission data from both \fermi/GBM and \kw observatories to characterize the burst's emission behavior. For time-integrated spectral analysis, we relied on \kw data, as it provided continuous coverage throughout the entire duration of the burst, enabling broadband characterization from the keV to MeV range (see Table \ref{tab:kw_spectral}). For the episode-wise spectral analysis, we employed \kw data for both emission episodes, while spectra from \fermi/GBM were used for the first episode (see Tables \ref{tab:DIC} and \ref{tab:bestfitparameters}), providing complementary low-energy sensitivity. 
Spectral fitting of the \kw data was performed with \texttt{XSPEC} (version 12.15.0), using the $\chi^2$ minimization technique. To ensure the validity of the $\chi^2$ statistic, energy channels were binned to achieve a minimum of 20 counts per spectral bin. 

The relatively high signal-to-noise ratio of GRB~220107A allowed for detailed time-resolved spectroscopy (see Tables \ref{tab:bestfitparameters} and \ref{tab:kw_time_resolved}). For \kw, finer temporal bins were defined based on fixed accumulation times to capture the burst's spectral evolution. For \fermi/GBM, temporal binning was determined using the Bayesian blocks algorithm, which adaptively identifies statistically significant changes in count rate. This approach yielded a set of four statistically significant time bins (e.g., as summarized in Table~\ref{tab:DIC}), which effectively traced the burst's temporal evolution. Spectral files for each time bin were generated with background levels estimated from quiescent intervals immediately preceding and following the burst's main emission. The subsequent fitting of \fermi/GBM time-resolved spectra was performed within the Multi-Mission Maximum Likelihood (\texttt{3ML}; \citealt{Vianello_etal_2015}) framework. The analysis excluded energy ranges known to be affected by instrumental features \citep{2009ExA....24...47B}, such as the iodine K-edge present in GBM NaI detectors\footnote{\url{https://fermi.gsfc.nasa.gov/ssc/data/analysis/GBM_caveats.html}}. All spectral parameters reported in this work are given with $1\sigma$ ($68\%$) confidence intervals, unless otherwise specified. For the \fermi/GBM analysis, model selection was conducted by the Deviance Information Criterion (DIC; \citealt{1978AnSta...6..461S}), prioritizing models that achieved the best fit with the lowest DIC values to ensure robust statistical inference.

\subsection{Polarimetric data analysis}
\label{polan}

Spectral analysis of GRBs alone can often lead to ambiguities in the selection of the most appropriate emission model. Polarization measurements provide a critical complementary observable to resolve these ambiguities. The polarization of X-rays and gamma-rays is fundamentally tied to their interaction with matter, where the differential cross-section for processes like Compton scattering depends on the polarization orientation, modulating the intensity of scattered photons or electrons \citep{2017NewAR..76....1M}. The \astrosat/CZTI is specifically designed to perform polarimetric measurements in the hard X-ray energy range (above 100~keV), leveraging its pixelated detector array to function as a Compton polarimeter. By detecting coincident events in adjacent pixels within a narrow temporal coincidence window of 20~$\mu$s and applying energy ratio criteria (in the range of 1:6) to filter out non-Compton events \citep{2022ApJ...936...12C}, CZTI enhances signal purity while minimizing background contamination. This capability enables the instrument to analyze the azimuthal distribution of Compton-scattered events, which exhibits modulation dependent on the polarization state of the incident radiation \citep{2021JApA...42..106C}. Through this analysis, CZTI can determine the polarization fraction and angle of GRBs, providing insights into the emission geometry and magnetic field structure of these energetic transients.

The CZTI polarization analysis for GRB~220107A follows the standard pipeline developed for GRB polarimetry with \astrosat/CZTI (e.g., \citealt{2022ApJ...936...12C, 2024ApJ...972..166G}). We briefly summarize the procedure below:

\begin{itemize}

\item {Selection of Compton events:} Compton events are identified as double-pixel interactions occurring within a 20~$\mu$s coincidence window. To eliminate spurious double events caused by random coincidences, we apply Compton scattering criteria, evaluating the energy deposition ratio between adjacent pixels.

\item {Generation of background-subtracted azimuthal angle distribution:} Compton events are extracted from both the GRB emission time window and the pre- and post-burst background intervals, excluding times affected by SAA passages.

\item {Correction for geometric and off-axis effects:} The background-subtracted azimuthal scattering angle distribution is then corrected for geometric asymmetries and off-axis response using Geant4 simulations of the \astrosat mass model \citep{GEANT4:2002zbu,2019ApJ_Chattopadhyay,2021JApA...42...93M}. The simulations incorporate the GRB spectral parameters and source direction to generate the expected distribution for unpolarized radiation, which is used to normalize the observed distribution.

\item {Determination of modulation amplitude and PA:} The corrected azimuthal angle distribution is modeled using a sinusoidal function (equation~\ref{eq:cosine_equation}), where the modulation factor is defined as $\mu = A/B$ and the polarization angle corresponds to $\phi_0$ within the CZTI detector plane. The fitting is carried out with the Markov Chain Monte Carlo (MCMC) method implemented in PyMC3\footnote{\url{https://pypi.org/project/pymc3/}}, ensuring robust parameter estimation and a proper treatment of statistical uncertainties.

\begin{equation}
C(\phi) = A \cos\left[2\left(\phi - \phi_0 + \frac{\pi}{2}\right)\right] + B
\label{eq:cosine_equation}
\end{equation}

\item {Estimation of PF:} 
The polarization fraction is obtained by normalizing the observed modulation factor ($\mu$) with the modulation amplitude for fully polarized radiation ($\mu_{100}$), obtained through Geant4 simulations tailored to the GRB's spectral parameters and direction relative to the \astrosat mass model. After deriving the posterior distributions of PF and PA from the MCMC fitting, we discard unphysical samples with PF $> 100\%$. From the filtered posterior, PA and PF are reported for statistically significant polarization detection utilizing the Bayes Factor (BF). According to Jeffreys' scale \citep{jeffreys1961, Kass_Raftery1995}, a BF $> 3.2$ provides positive evidence in favor of the alternative hypothesis---namely, the sinusoidal model, which explains the presence of polarization. The null hypothesis corresponds to a linear model, described by a straight line, which represents the unpolarized case and could indicate a uniform distribution of Compton events across all eight azimuthal scattering angles in the detector plane. A BF less than 1 supports the null hypothesis. For the time-integrated emission, we report 2$\sigma$ upper limits to facilitate comparison with previous GRB polarimetry results (e.g., POLAR), which conventionally quote time-integrated constraints at this confidence level. For the time-resolved analysis, the first emission episode contains a comparatively small number of valid Compton events, resulting in a weakly constrained modulation amplitude. Therefore, we report a $1.5\sigma$ upper limit on the PF for this episode to provide a meaningful constraint. The second episode has higher Compton statistics, enabling a robust $2\sigma$ upper limit. This upper limit is determined using the method described in \cite{2022ApJ...936...12C}. 

\end{itemize}

\begin{figure*}
    \centering
    \includegraphics[scale=0.35]{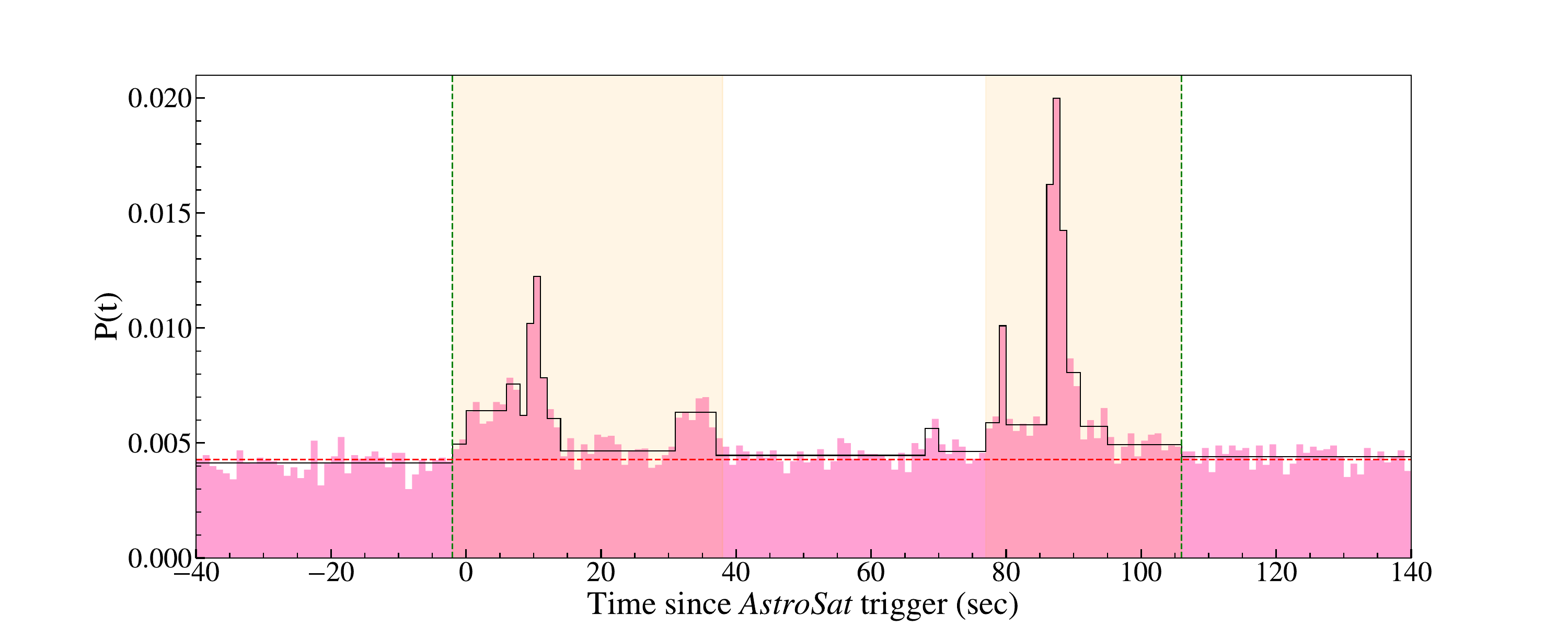}
    \caption{\astrosat/CZTI Compton light curve of GRB 220107A, illustrating the temporal distribution of photon counts as a function of time since the \fermiT. Bayesian block analysis is employed to delineate the time-integrated duration, as well as the specific intervals for episode 1 and episode 2, which are selected for spectro-polarimetric analysis. The vertical green lines mark the start and end times of these intervals, while the shaded beige regions highlight the durations of the first and second episodes, respectively. The red dashed line represents the baseline count rate, with peak intensities observed around -2-38 seconds and 77-106 seconds post-trigger, indicating significant emission episodes.}
    \label{fig:cztilc}
\end{figure*}

\section{Results}
\label{sec:results}

\subsection{Spectral measurement for GRB 220107A}

\begin{table*}[ht]
\centering
\caption{Spectral fit results of GRB 220107A from \kw data. Results are shown for the time-averaged prompt emission (0--98 sec), episode 1 (0--33 sec), and episode 2 (66--98 sec) of \thisgrb. Quoted uncertainties correspond to the 1 $\sigma$ confidence level.}
\label{tab:kw_spectral}
\begin{tabular}{lcccccccccc}
\hline\hline
Interval & Model & $\alpha$ & $\beta$ & $E_{\rm p}$ (keV) & $kT$ (keV) & Flux$^{a}$ & $\chi^{2}$/dof \\
\hline
Time-integrated (0--98 sec) & Band     & $-1.01 \pm 0.07$ & $-3.07^{+0.43}_{\rm unconstrained}$ & $227^{+17}_{-16}$ & --    & $5.53^{+0.58}_{-0.55}$ & 101.0/97 \\
                          & CPL      & $-1.04 \pm 0.06$ & --                        & $235^{+14}_{-12}$ & --    & $5.07^{+0.18}_{-0.17}$ & 101.6/98 \\
                          & Band+BB  & $-0.96^{+0.22}_{-0.14}$ & $-3.11^{+0.46}_{\rm unconstrained}$ & $234^{+24}_{-18}$ & $14.8^{\rm unconstrained}_{-4.5}$ & $5.39^{+0.62}_{-0.57}$ & 100.6/95 \\
                          & CPL+BB   & $-1.00 \pm 0.15$ & --                        & $243^{+22}_{-17}$ & $16.9^{+21.7}_{-16.6}$ & $4.98^{+0.22}_{-0.26}$ & 101.0/96 \\
\hline
Episode 1 (0--33 sec)         & Band     & $-0.96 \pm 0.09$ & $<-3.55$                  & $260^{+13}_{-16}$ & --    & $5.31^{+0.32}_{-0.13}$ & 133.2/97 \\
                          & CPL      & $-0.96 \pm 0.09$ & --                        & $260^{+23}_{-19}$ & --    & $5.31^{+0.28}_{-0.26}$ & 133.2/98 \\
                          & Band+BB  & $-0.48 \pm 0.37$ & $<-3.57$                  & $270^{+23}_{-19}$ & $13.7 \pm 2.1$ & $4.61^{+0.54}_{-0.57}$ & 130.7/95 \\
                          & CPL+BB   & $-0.47^{+0.59}_{-0.38}$ & --                   & $270^{+23}_{-19}$ & $13.7 \pm 2.1$ & $4.60^{+0.55}_{-0.56}$ & 130.0/96 \\
\hline
Episode 2 (66--98 sec)        & Band     & $-0.97 \pm 0.07$ & $-2.68^{+0.17}_{-0.25}$   & $195^{+11}_{-10}$ & --    & $10.3^{+0.8}_{-0.7}$ & 106.3/97 \\
                          & CPL      & $-1.03 \pm 0.06$ & --                        & $210 \pm 11$      & --    & $8.66^{+0.27}_{-0.26}$ & 112.2/98 \\
                          & Band+BB  & $-0.72^{+0.29}_{-0.20}$ & $-2.68^{+0.17}_{-0.23}$ & $189 \pm 11$ & $6.97^{+1.96}_{-3.05}$ & $9.87^{+0.82}_{-0.76}$ & 103.8/95 \\
                          & CPL+BB   & $-1.45^{+0.18}_{-0.16}$ & --                    & $275^{+94}_{-48}$ & $41.8 \pm 3.6$ & $7.87^{+0.53}_{-0.51}$ & 106.6/96 \\
\hline
\end{tabular}
\begin{flushleft}
\footnotesize
$^{a}$Fluxes are quoted in units of $10^{-7}$ erg cm$^{-2}$ sec$^{-1}$ over 10 keV–10 MeV.  
\end{flushleft}
\end{table*}

\begin{table*}
\centering
\caption{The DIC values obtained from the episode-1 analysis and from the individual bins within this episode using \fermi/GBM data. The model with the lowest DIC (highlighted in bold) is considered the best fit.}
\label{tab:DIC}
\begin{tabular}{l|cccc}\hline				
{Time-intervals}& & &{DIC value} &\\
(from \fermiT in sec)&{Band}&{CPL}&{BB+Band}&{BB+CPL}\\  
\hline 
-2.00-38.00& 5807.8& 5814.4&  5772.6 & {\bf 5732.2}\\\hline
-2.00-6.40& 4162.03& 4164.59&4118.32 & {\bf 4058.43} \\
6.40-10.01&3327.57 &3336.69 & {\bf 3285.81} & 3313.24\\
10.01-11.24& 2306.39  & 2309.21 & 2288.72 & {\bf 2287.70}\\
11.24-38.00& 5343.32 & 5359.53 & 5292.22 & {\bf 5202.4} \\
\hline
\end{tabular}
\end{table*}

\begin{table*}
\centering
\caption{The best-fit spectral parameters for the first episode and the individual bins within this episode using \fermi/GBM data. Errors are quoted at the 1$\sigma$ confidence level.}
\label{tab:bestfitparameters}
\begin{tabular}{l|cccccc}\hline				
{Time-intervals}& & & &Best-fit parameters & &\\
(from \fermiT in sec)& $\it \alpha_{\rm pt}$&\Ep (keV)& $\it \beta_{\rm pt}$& $\it \Gamma_{\rm CPL}$& $E_{\rm c}$ (keV)& $kT$ (keV)\\  
\hline 
-2.00-38.00& -- & -- & -- & -0.62$^{+0.09}_{-0.12}$ & 183.22$^{+43.65}_{-30.08}$ &13.98$^{+4.23}_{-2.41}$\\
\hline
-2.00-6.40 & -- &-- & -- & $-0.61^{+0.17}_{-0.14}$  & $111.22^{+40.79}_{-5.72}$  & 10.13$^{+5.80}_{-1.02}$  \\
6.40-10.01 & $-0.61^{+0.01}_{-0.18}$ &$212.40^{+48.27}_{-48.27}$ & -2.10$^{+0.29}_{-0.29}$ & --  & --  & 10.48$^{+16.44}_{-0.56}$  \\
10.01-11.24 & -- &-- & -- & $-0.59^{+0.09}_{-0.05}$  & $266.37^{+32.95}_{-28.09}$  & 9.41$^{+5.99}_{-0.71}$  \\
11.24-38.00 & -- &-- & -- & $-0.52^{+0.12}_{-0.26}$  & $119.20^{+63.62}_{-63.62}$  & 11.31$^{+9.92}_{\rm unconstrained}$  \\
\hline
\end{tabular}
\end{table*}

\subsubsection{Time-averaged and episode-wise spectral analysis}
\label{spectra}

We applied the Bayesian block algorithm \citep{scargle2013studies} with a false alarm probability of $p_0 = 0.01$ to the CZTI Compton light curves in order to identify statistically significant changes, which were then used to define the intervals for spectral and polarization analyses of GRB 220107A (see Figure \ref{fig:cztilc}). The time bins corresponding to the two main emission episodes (hereafter episode 1 and episode 2) were extracted, along with a full time-integrated window. For the \kw spectral analysis, we selected time windows that closely matched the Bayesian block intervals obtained from \astrosat/CZTI Compton light curves, ensuring consistency between the spectral and polarimetric studies (see Section~\ref{sec:polarization}). 

The time-averaged \kw spectrum (\fermiT+0 to \fermiT+98 sec) was fitted with several spectral models, including a cutoff power law (CPL), the Band function, and their extensions with a thermal blackbody (BB) component (Band+BB, and CPL+BB). The Band and CPL functions provided statistically acceptable fits (comparison based on reduced $\chi^2$ values), with $\chi^{2}/{\rm d.o.f.} = 101.0/97$ and $101.6/98$, respectively. Adding a thermal component (yielding Band+BB or CPL+BB) did not improve the fit compared to their simpler counterparts, with $\chi^{2}/{\rm d.o.f.} = 100.6/95$ and $101.0/96$. An $F$-test further evaluated on these nested models corroborated the lack of substantial improvement from the BB addition, with $F = 0.19$ ($p = 0.83$) for Band+BB and $F = 0.29$ ($p = 0.75$) for CPL+BB. Consequently, the simpler Band model is preferred due to its adequate fit to the time-averaged spectrum of \thisgrb. The fit parameters are summarized in Table~\ref{tab:kw_spectral}.

For the first episode (\fermiT+0 to \fermiT+33 sec) of \thisgrb, the Band and CPL models gave nearly identical fits with $\chi^{2}/{\rm d.o.f.} = 133.2/97$ and $133.2/98$, respectively, indicating that the additional high-energy power-law tail ($\beta$) in the Band function is unconstrained. When an additional blackbody component is introduced, the fit improved slightly to $\chi^{2}/{\rm d.o.f.} = 130.7/95$ (Band+BB) and $130.0/96$ (CPL+BB). The thermal component is characterized by $kT \approx 13.7$ keV, supporting the presence of a photospheric emission in the first episode. Additionally, the $\alpha$ value of Band+BB or CPL+BB remains hard, consistent with the photospheric emission model. Overall, the CPL+BB model is statistically preferred for episode 1 (see Table \ref{tab:kw_spectral}). This supports the interpretation that the early emission episode carries a more prominent quasi-thermal/photospheric signature.

\begin{figure}[!ht]
\centering
\includegraphics[scale=0.45]{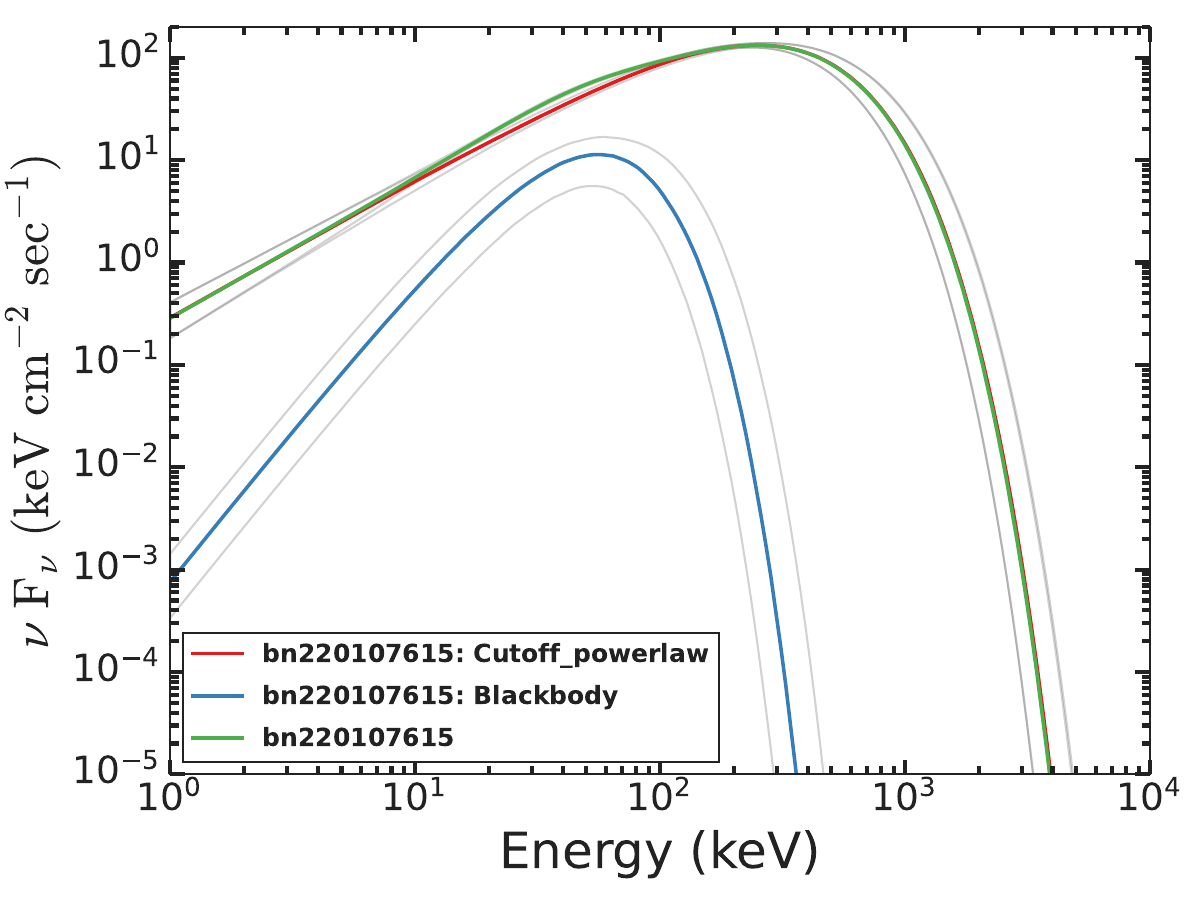}
\caption{Time-integrated $\nu F_{\nu}$ spectrum of GRB 220107A (episode 1; $T_{0}-2$ to $T_{0}+38$ sec) fitted with the cutoff power law plus blackbody (CPL+BB) model using \fermi/GBM data. The red curve shows the cutoff power law component, the blue curve shows the blackbody emission, and the green curve represents the total model. The shaded gray regions denote the $1\sigma$ uncertainties of each spectral component. The fit suggests a thermal blackbody component at $kT \sim 14$ keV in addition to the non-thermal CPL continuum, which significantly improves the fit relative to purely non-thermal models.}
\label{fig:gbmspectrum}
\end{figure}

\begin{table*}[t]
\centering
\caption{Time-resolved spectral fit results of GRB~220107A with \kw data. Errors are quoted at the 1$\sigma$ confidence level.}
\label{tab:kw_time_resolved}
\begin{tabular}{cccccccccc}
\hline\hline
Interval & Time (s) & Model & $\alpha$ & $\beta$ & $E_{\rm p}$ (keV) & $kT$ (keV) & Flux ($\rm erg\,cm^{-2}\,s^{-1}$) & $\chi^{2}$/dof \\
\hline
sp1--4   & 0.0--0.256 & CPL     & $-0.49^{+0.45}_{-0.36}$ & -- & $245^{+62}_{-45}$ & -- & $(1.71^{+0.30}_{-0.27})\times10^{-6}$ & 12.1/32 \\
         &            & CPL+BB  & $-0.20^{+0.20}_{-0.62}$ & -- & $246^{+97}_{-45}$ & $10.5$ (n/c) & $(1.62^{+0.37}_{-0.78})\times10^{-6}$ & 11.9/30 \\
\hline
sp5      & 0.256--8.448 & CPL    & $-0.74\pm0.10$ & -- & $291^{+22}_{-19}$ & -- & $(1.18\pm0.06)\times10^{-6}$ & 112.6/98 \\
         &              & CPL+BB & $0.09^{+0.57}_{-0.44}$ & -- & $278^{+18}_{-16}$ & $12.7^{+1.8}_{-2.0}$ & $(9.91^{+0.97}_{-0.91})\times10^{-7}$ & 107.4/96 \\
\hline
sp8      & 24.8--33.0  & CPL     & $-0.81\pm0.16$ & -- & $174^{+20}_{-16}$ & -- & $(5.8\pm0.4)\times10^{-7}$ & 85.7/98 \\
         &             & CPL+BB  & $-0.29^{+0.96}_{-0.51}$ & -- & $243^{+46}_{-36}$ & $17.6^{+3.1}_{-2.2}$ & $(4.55^{+0.74}_{-0.84})\times10^{-7}$ & 78.4/96 \\
\hline
sp11     & 49.4--57.6  & CPL     & $-1.32^{+0.38}_{-0.29}$ & -- & $428^{+559}_{-168}$ & -- & $(2.37^{+1.01}_{-0.57})\times10^{-7}$ & -- \\
\hline
sp13     & 65.8--74.0  & CPL     & $-1.18\pm0.18$ & -- & $330^{+119}_{-70}$ & -- & $(5.3^{+0.9}_{-0.7})\times10^{-7}$ & 94.8/98 \\
\hline
sp14     & 74.0--81.4  & 
        Band    & $-0.81\pm0.10$ & $-3.06^{+0.34}_{-1.28}$ & $191^{+13}_{-12}$ & -- & $(1.70\pm0.15)\times10^{-6}$ & 78.2/97 \\
\hline
sp15     & 81.4--89.6  & 
          Band    & $-1.07\pm0.10$ & $-2.95^{+0.26}_{-0.79}$ & $188^{+18}_{-14}$ & -- & $(1.38\pm0.15)\times10^{-6}$ & 90.1/97 \\
\hline
sp14--15 & 74.0--89.6  & 
 Band    & $-0.93\pm0.07$ & -- & $189^{+10}_{-9}$ & -- & $(1.57\pm0.10)\times10^{-6}$ & 85.9/97 \\
\hline
sp16     & 89.6--97.8  & CPL     & $-1.40^{+0.49}_{-0.40}$ & -- & $182^{+551}_{-58}$ & -- & $(2.01^{+1.31}_{-0.41})\times10^{-7}$ & 92.8/98 \\
\hline
\end{tabular}
\end{table*}

We also performed time-integrated spectral analysis of the first episode of GRB 220107A in the interval from \fermiT-2 to \fermiT+38 s using \fermi/GBM data. Model comparison based on the DIC (see Table {\ref{tab:DIC}}) shows that the CPL+BB model ($\mathrm{DIC}=5732.2$) provides a significantly better description of the spectrum compared to Band ($\mathrm{DIC}=5807.8$), CPL ($\mathrm{DIC}=5814.4$), or Band+BB ($\mathrm{DIC}=5772.6$). The best-fitting CPL+BB model yields a low-energy photon index of $\alpha=-0.62^{+0.10}_{-0.12}$, a cutoff energy of $E_{\mathrm{c}}=183^{+44}_{-30}$ keV, and a normalization of $K_{\mathrm{CPL}}=0.26^{+0.15}_{-0.07}$ keV$^{-1}$ cm$^{-2}$ sec$^{-1}$ (see Table {\ref{tab:bestfitparameters}}). The additional thermal component is characterized by a blackbody temperature of $kT=13.98^{+4.23}_{-2.41}$ keV. The statistical improvement and well-constrained parameters indicate the presence of a sub-dominant but significant thermal emission component (see Figure \ref{fig:gbmspectrum}).

In contrast, the second episode (\fermiT+66 to \fermiT+98 sec) analyzed using \kw data was dominated by non-thermal radiation. The Band function provides a statistically superior fit compared to the CPL model ($\chi^{2}/\mathrm{dof} = 106.3/97$ versus $112.2/98$), with an F-test indicating that the inclusion of a high-energy spectral index is required ($p \approx 0.02$). The addition of a blackbody component to either model further reduces the fit statistic (Band+BB: $103.8/95$; CPL+BB: $106.6/96$), but the improvements are not statistically significant for Band+BB ($p \approx 0.32$) and are only marginal for CPL+BB ($p \approx 0.09$). Additionally, the $\alpha$ value of Band+BB or CPL+BB remains soft (see Table \ref{tab:kw_spectral}), consistent with the synchrotron emission model. Therefore, the spectrum of episode~2 is best described by the Band function without the need for an additional thermal component.

\subsubsection{Spectral parameter evolution}

\begin{figure*}
\centering
\includegraphics[scale=.5]{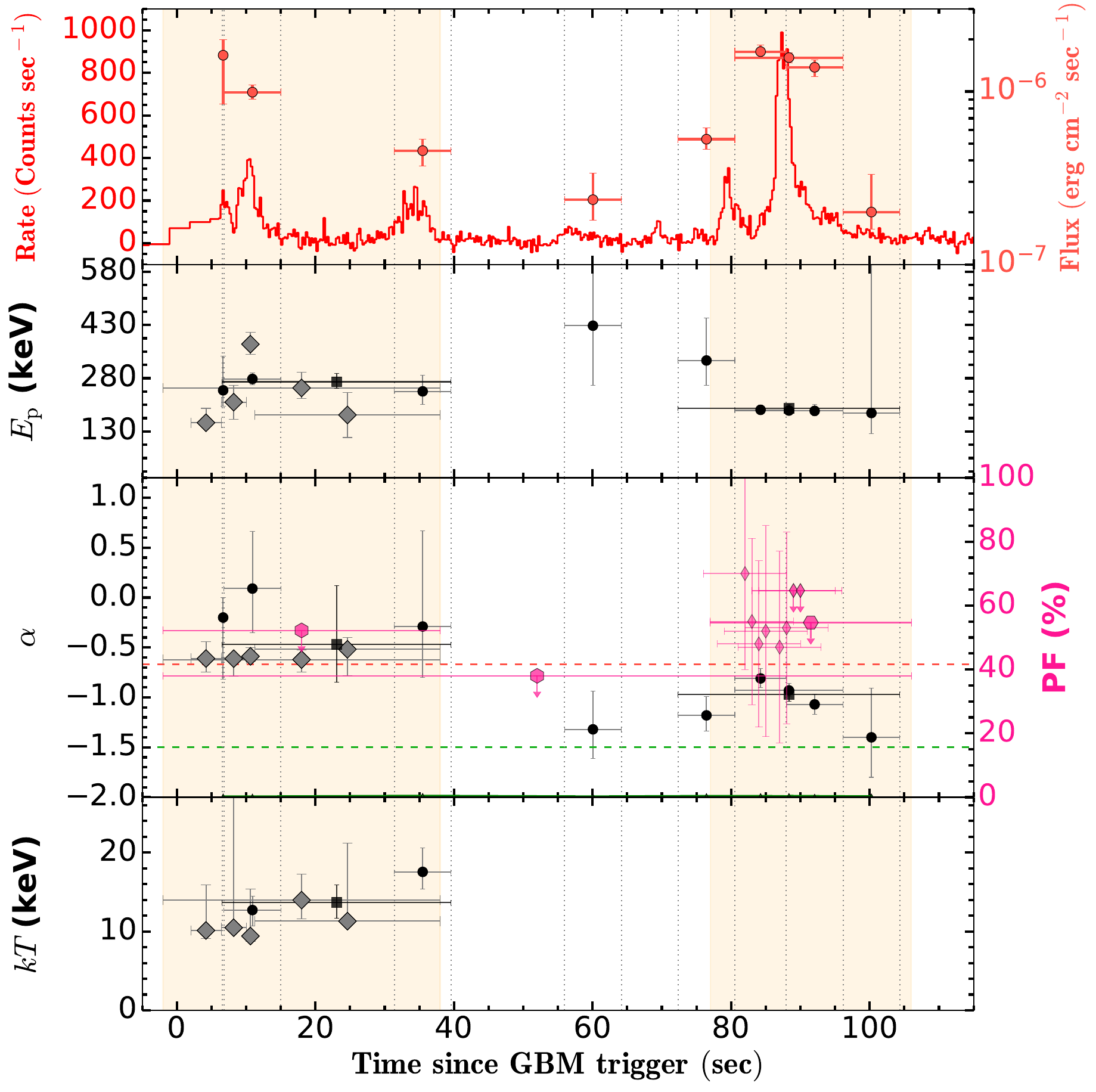}
\caption{Spectro-polarimetric evolution of GRB 220107A: Time-resolved spectro-polarimetric analysis of GRB 220107A. {Top panel:} \kw light curve in the 20–1300~keV band (red) with flux evolution (orange-red) shown on the right axis. Shaded regions indicate the two main episodes of the burst. Vertical gray dashed lines mark finer time bins used for spectral analysis using \kw data. {Second panel:} Evolution of $E_{\rm p}$ with time, showing the peak energy of the emission in each time bin. {Third panel:} Evolution of the low-energy spectral index $\alpha$ with synchrotron slow- ($\alpha=-2/3$) and fast-cooling ($\alpha=-3/2$) reference lines indicated by dashed lines. Pink points indicate the polarization fraction measured with \astrosat~CZTI, including upper limits. {Bottom panel:} Evolution of the blackbody temperature $kT$ for models including a thermal component. The combination of spectral and polarimetric measurements highlights the hard-to-soft evolution of the first to second episode and provides constraints on the emission mechanisms of GRB~220107A.}
    \label{fig:TRS}
\end{figure*}

We performed a detailed time-resolved spectral analysis of GRB~220107A using \kw data, fitting the spectra with single-component models (CPL, Band function) as well as combinations with a blackbody. To observe spectral evolution within each episode, finer time bins (shown as vertical gray dashed lines in Figure \ref{fig:TRS}) were selected based on photon accumulation in \kw to study the detailed evolution of the spectral parameters. The time-resolved spectral analysis results are summarized in Table~\ref{tab:kw_time_resolved}. 

During the initial episode (sp1--4, sp5, and sp8; see Table \ref{tab:kw_time_resolved} for the interval names), the CPL model provides a reasonable fit with $E_{\rm p} \sim 174$--$291$~keV. The addition of a thermal component (CPL+BB) marginally improves the fit ($\Delta\chi^{2} \sim 2$), yielding blackbody temperatures of $kT \approx 10$--18~keV, consistent with photospheric emission.  In the later intervals (sp11--13), the spectra remain adequately described by a CPL with $E_{\rm p}$ evolving from $\sim 330$~keV to $\sim 428$~keV. The brightest phases of the second emission episode (sp14--16) are best fit with the Band or CPL function. 

The evolution of $E_{\rm p}$, $\alpha$, and $kT$ for best fit models is shown in the three lower panels of Figure \ref{fig:TRS}. The spectral evolution suggests compelling behavior: during the first episode, the low-energy photon index $\alpha$ exhibits hard values, significantly harder than the typical synchrotron limit of $-2/3$ \citep{1998ApJ...506L..23P, 2000ApJS..126...19P}, which is not unusual. This initial hard phase is followed by a pronounced softening during the second episode (brighter episode), where $\alpha$ approaches values consistent with synchrotron emission mechanisms \citep{1994ApJ...432L.107K, 2018MNRAS.476.1785B, 2020NatAs...4..210Z, 2020NatAs...4..174B}. Synchrotron slow- and fast-cooling lines are indicated with dashed lines in Figure \ref{fig:TRS} to guide comparison with theoretical expectations. The peak energy $E_\mathrm{p}$ also evolves, following intensity tracing behavior in episode~1 and episode~2, though with larger uncertainties. The thermal component $kT$ is detected in the CPL+BB model and shows moderate evolution during the first episode of the burst.

\subsection{Redshift measurement and Prompt correlation}
\label{AMATI}

\begin{figure*}[!ht]
    \centering
    \includegraphics[scale=0.27]{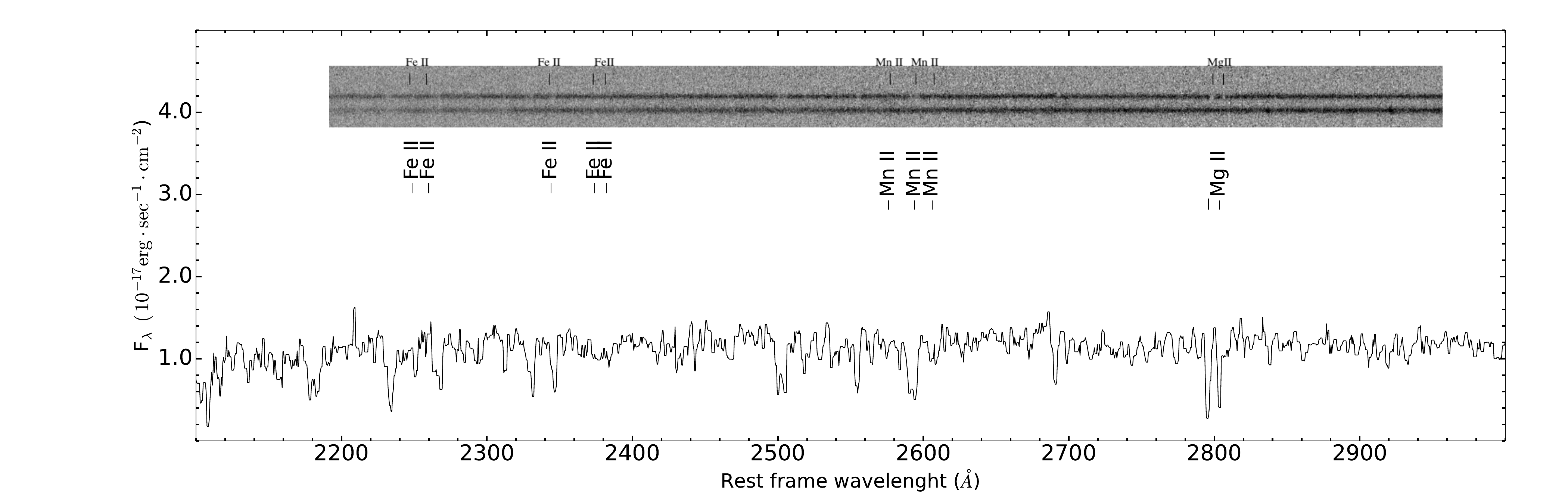}
    \caption{BTA/SCORPIO-II spectrum of the optical afterglow of GRB~220107A obtained 1.36~days post-burst. Top panel: The two-dimensional spectrum showing the trace of the GRB afterglow (upper) and an unrelated foreground galaxy at $z = 0.309$ (lower, \citealt{2022GCN.31419....1P, 2022GCN.31423....1C}). Bottom panel: The flux-calibrated one-dimensional spectrum smoothed with a $4\times4$ pixel median filter. Prominent absorption features, including \ion{Fe}{II}, \ion{Mn}{II}, and the \ion{Mg}{II} doublet, are marked. All identified lines are consistent with a common absorption system at $z=1.246$, which we consider as the redshift of the GRB.}
    \label{fig:redhift}
\end{figure*}

\begin{figure*}[!ht]
    \centering
    \includegraphics[scale=.5]{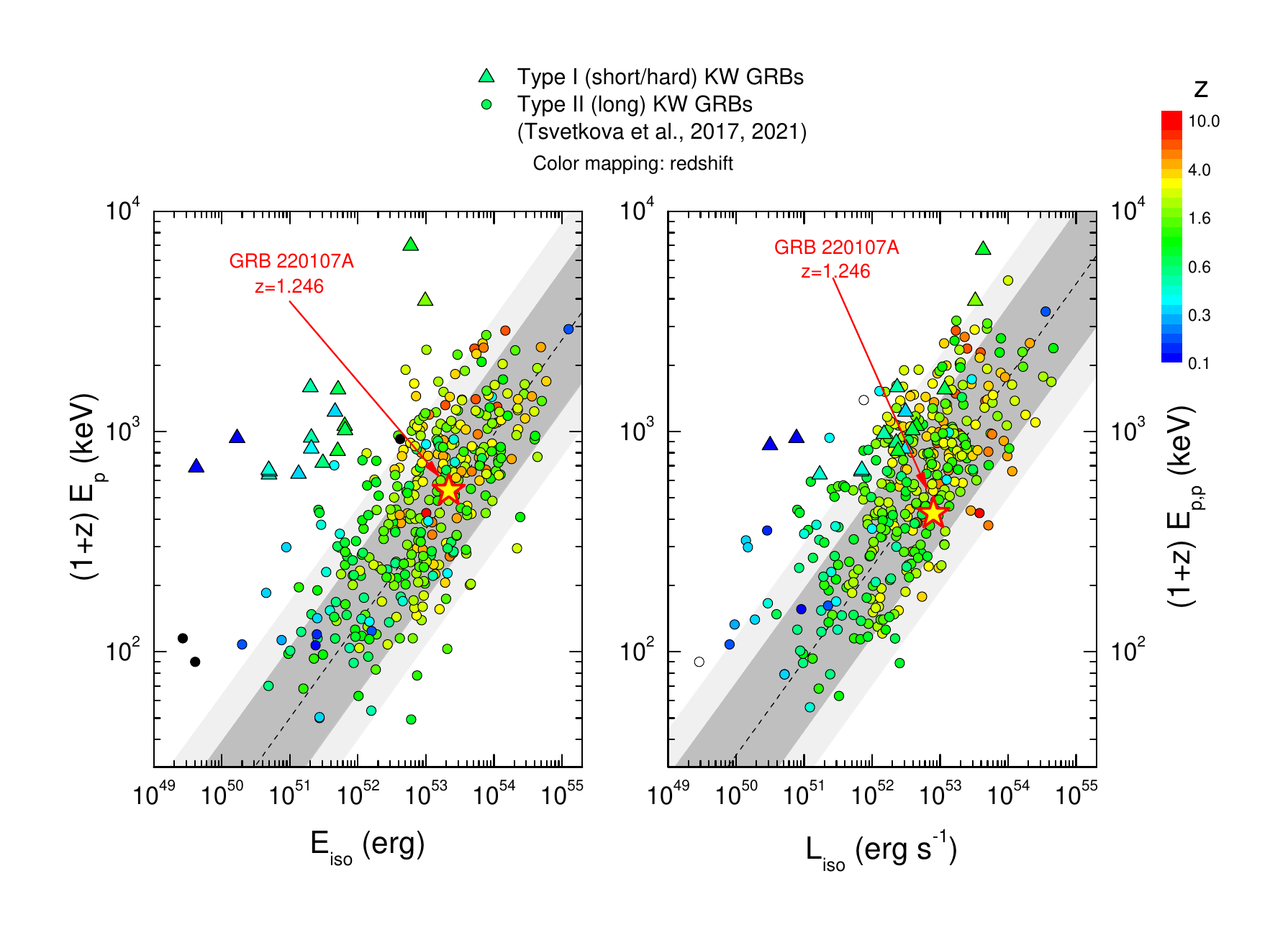}
    \caption{GRB~220107A (red star) in the Amati (left) and Yonetoku (right) planes based on \kw data. The sample of Type~I (short/hard; triangles) and Type~II (long; circles) GRBs is taken from \cite{2017ApJ...850..161T, 2021ApJ...908...83T}, with points color-coded by redshift. The shaded regions denote the 1$\sigma$ and 2$\sigma$ scatter around the best-fit correlations. For GRB~220107A, we derive $E_{\rm iso}=(2.15\pm0.08)\times10^{53}$ erg and $(1+z)E_{\rm p,i}=(540\pm34)$ keV from the time-integrated spectrum, and $L_{\rm iso}=(7.86\pm0.57)\times10^{52}$ erg s$^{-1}$ with $(1+z)E_{\rm p,p}=(425\pm35)$ keV from the peak spectrum. Both measurements confirm that GRB~220107A lies in good agreement with the established Amati and Yonetoku relations.}
    \label{fig:correlation}
\end{figure*}

Three spectra (600~s each, covering the 3700–8500~\AA{} range) of the optical afterglow of GRB~220107A were obtained using the 6\,m BTA telescope with the SCORPIO-II focal reducer at the Special Astrophysical Observatory on 2022 January 8, 23:25~UT (mid-exposure time, 1.36 days after the burst). The data were processed using standard procedures, including bias subtraction, flat-field correction, wavelength calibration, and flux calibration with spectrophotometric standard stars. The one-dimensional spectrum was extracted and smoothed with a $4\times4$ pixel median filter to enhance the signal-to-noise ratio. Several prominent absorption features are clearly detected in the rest-frame ultraviolet, including multiple \ion{Fe}{II} lines, the \ion{Mn}{II} triplet, and the \ion{Mg}{II} $\lambda\lambda$2796, 2803 doublet (see Figure \ref{fig:redhift}). All of these features are consistent with an absorption system at a common redshift of $z=1.246$ \citep{2022GCN.31423....1C}. While we do not detect excited fine-structure lines or host galaxy emission lines that would definitively associate this system with the GRB environment, we identify this as the redshift of GRB~220107A under the standard assumption that the highest-redshift absorption system corresponds to the host. Furthermore, the detection of the afterglow continuum down to the blue limit of our spectral coverage (3700~\AA{}) allows us to place a formal upper limit on the redshift. The absence of a Ly$\alpha$ break or significant Ly$\alpha$ forest absorption within our spectral window implies that the Ly$\alpha$ transition ($\lambda_{rest} = 1215.67$~\AA{}) must lie blueward of 3700~\AA{}. This constrains the redshift to $z < (3700/1215.67 - 1) \approx 2.04$. This limit is consistent with identifying the $z=1.246$ system as the GRB host \citep{2022GCN.31423....1C}. 

Several global correlations have been identified in the prompt emission properties of GRBs, and these relations are essential for understanding both their physical origin and potential use as cosmological probes \citep{2020MNRAS.492.1919M, 2025arXiv251016475A}. One of the best known is the Amati correlation, which connects the rest-frame spectral peak energy, $E_{\rm p,i}$, with the isotropic-equivalent radiated energy, $E_{\rm iso}$ \citep{2006MNRAS.372..233A}. For GRB~220107A, considering a redshift of $z=1.246$ and standard cosmology ($H_{0}=70$ km sec$^{-1}$ Mpc$^{-1}$, $\Omega_{M}=0.27$, $\Omega_{\Lambda}=0.73$), we obtained $E_{\rm iso}=(2.15\pm0.08)\times10^{53}$ erg and $(1+z)E_{\rm p,i}=(540\pm34)$ keV from the time-integrated \kw spectrum. These values place GRB~220107A well within the $E_{\rm p,i}$–$E_{\rm iso}$ distribution of long-duration GRBs, showing consistency with the Amati relation and supporting its classification as a Type II (long) event (see Figure \ref{fig:correlation}).

We also investigated the Yonetoku correlation, which connects the rest-frame spectral peak energy of the brightest emission episode with the peak isotropic luminosity, $L_{\rm iso}$ \citep{2010PASJ...62.1495Y}. From the peak spectrum (sp14–15), we derived $L_{\rm iso}=(7.86\pm0.57)\times10^{52}$ erg sec$^{-1}$ and $(1+z)E_{\rm p,p}=(425\pm35)$ keV. These parameters place GRB~220107A in close agreement with the best-fit Yonetoku relation, consistent with the behavior of the broader population of long GRBs (see Figure \ref{fig:correlation}). Taken together, the positions of GRB~220107A in both the Amati and Yonetoku planes indicate that its energetics and spectral properties are typical of the cosmological long-GRB population and follow the same empirical correlations that underpin much of our understanding of GRB prompt emission physics.

\subsection{Polarization measurement for GRB 220107A}
\label{sec:polarization}

\begin{figure}
\centering
\begin{overpic}[scale=0.35]{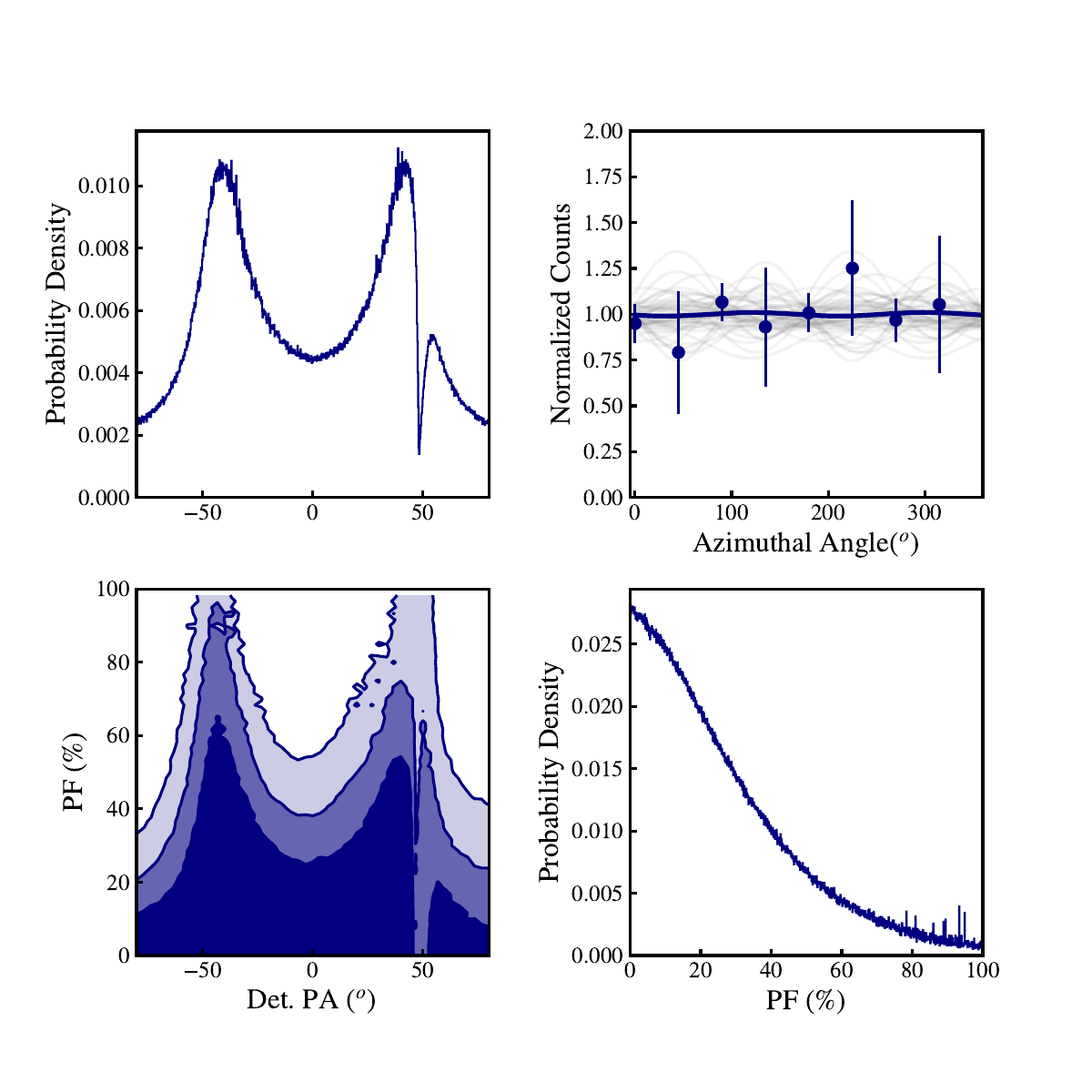}
        \put(5,90){{(a)}} 
    \end{overpic}
    \hfill
    \begin{overpic}[scale=0.35]{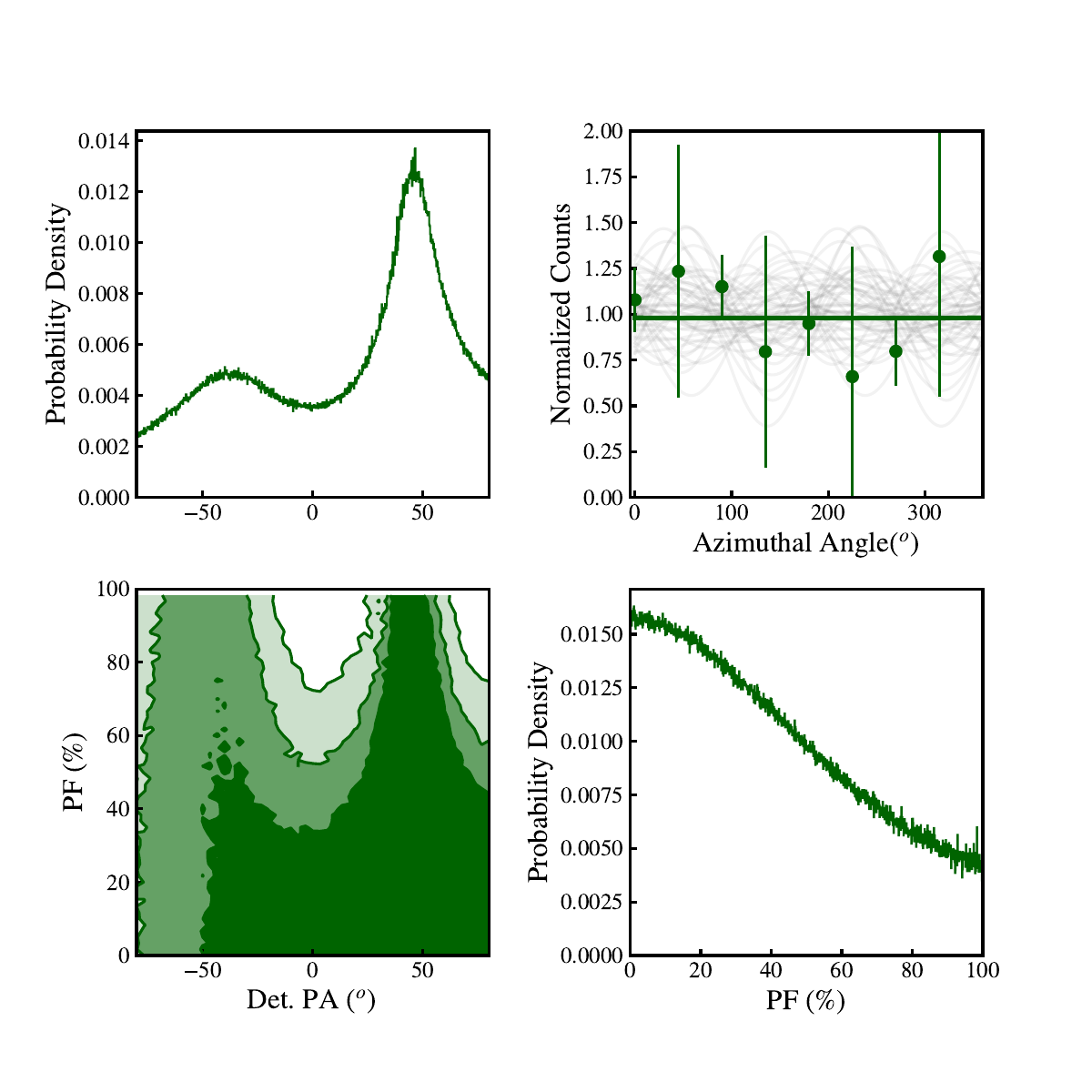}
        \put(5,90){{(b)}}
    \end{overpic}
    \hfill
    \begin{overpic}[scale=0.35]{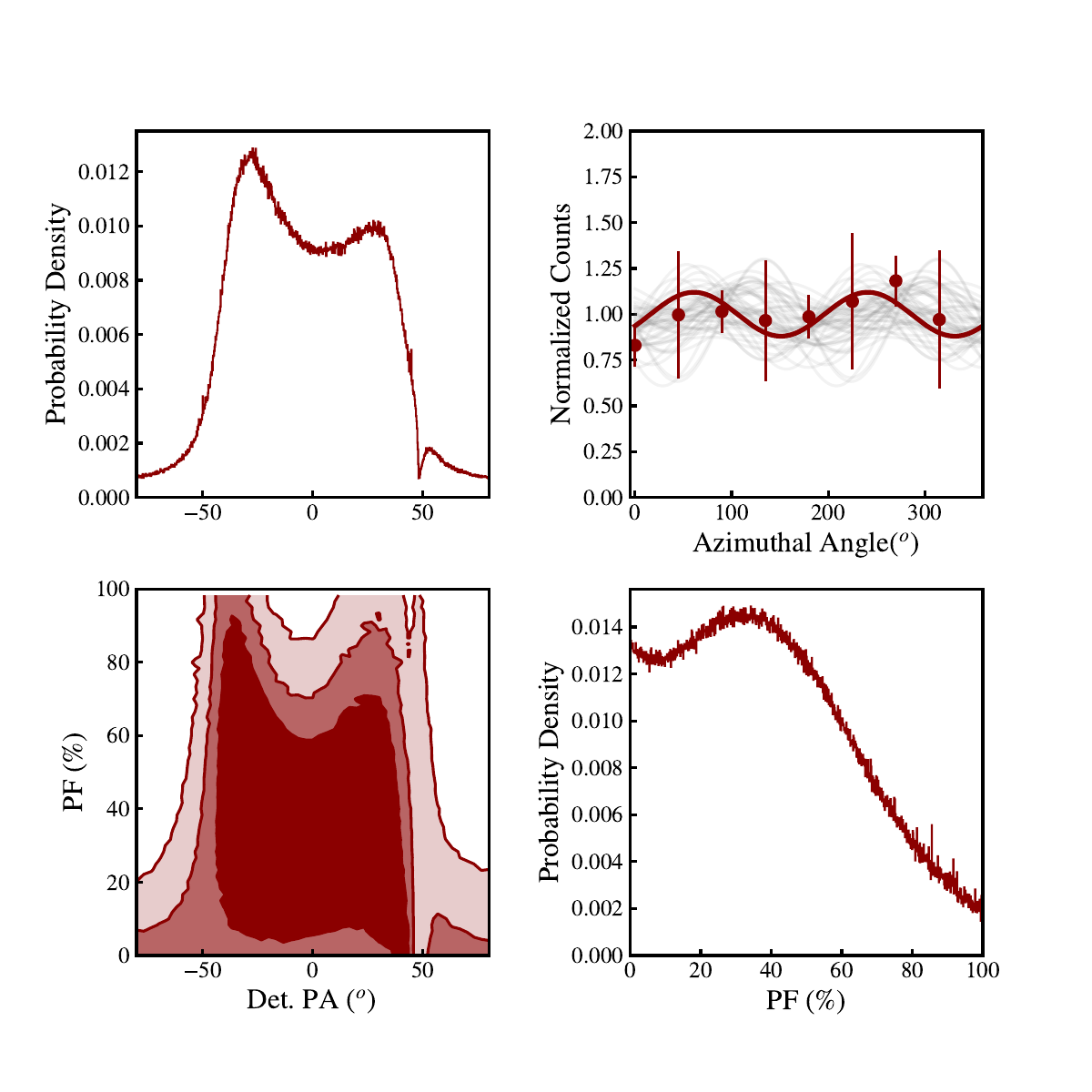}
        \put(5,90){{(c)}}
    \end{overpic}
    \caption{Each plot contains four subpanels: the top-left shows the posterior distribution of the polarization angle in the detector plane (Det. PA); the top-right shows the best-fit modulation curve (solid line) together with 1000 randomly selected realizations from the MCMC simulations (grey lines); the bottom-left shows the joint posterior distribution of PF and Det. PA with contours corresponding to the 68\%, 95\%, and 99\% credible regions; and the bottom-right shows the posterior distribution of the polarization fraction (PF). 
(a) Time-integrated case (\fermiT–2 to \fermiT+106 sec; contours in navy blue). 
(b) episode 1 (\fermiT–2 to \fermiT+38 sec; contours in green). 
(c) episode 2 (\fermiT+77 to \fermiT+106 sec; contours in red).}
\label{fig:PF}
\end{figure}

\subsubsection{Time-averaged polarization}

By integrating the light curve over the entire duration of the burst, the time-averaged analysis provides an overall picture of the polarization properties \citep{2022ApJ...936...12C}. This can reveal dominant emission components and average magnetic field configurations. However, it can mask significant temporal variations in polarization that might be indicative of changing emission mechanisms or outflow dynamics \citep{2011ApJ...743L..30Y, 2019A&A...627A.105B, 2019ApJ...882L..10S, 2020MNRAS.493.5218S, 2024ApJ...972..166G}. 

We performed a polarization analysis on the time-integrated prompt emission of GRB 220107A, covering the interval from \fermiT–2 to \fermiT+106 sec, during which approximately 1279 Compton scattering events were recorded in \astrosat/CZTI. For this GRB, the minimum detectable polarization (MDP) is $28.75\%$ at the $99\%$  confidence level, estimated using the standard CZTI MDP formalism described by \citet{2014ExA....37..555C}. The data show no significant detection of polarization, with a Bayes factor of ${\rm BF} = 0.67$, consistent with an unpolarized signal (see Figure \ref{fig:PF}). A 2$\sigma$ upper limit on the polarization fraction was derived, yielding ${\rm PF} < 38\%$. Such non-detections are common in time-averaged analyses of GRBs, as any intrinsic polarization signatures are likely diluted when integrating over multiple emission episodes \citep{2019ApJ...882L..10S, 2020A&A...644A.124K, 2022ApJ...936...12C, 2024ApJ...972..166G}.

\begin{figure}[!ht]
\centering
\includegraphics[width=\hsize]{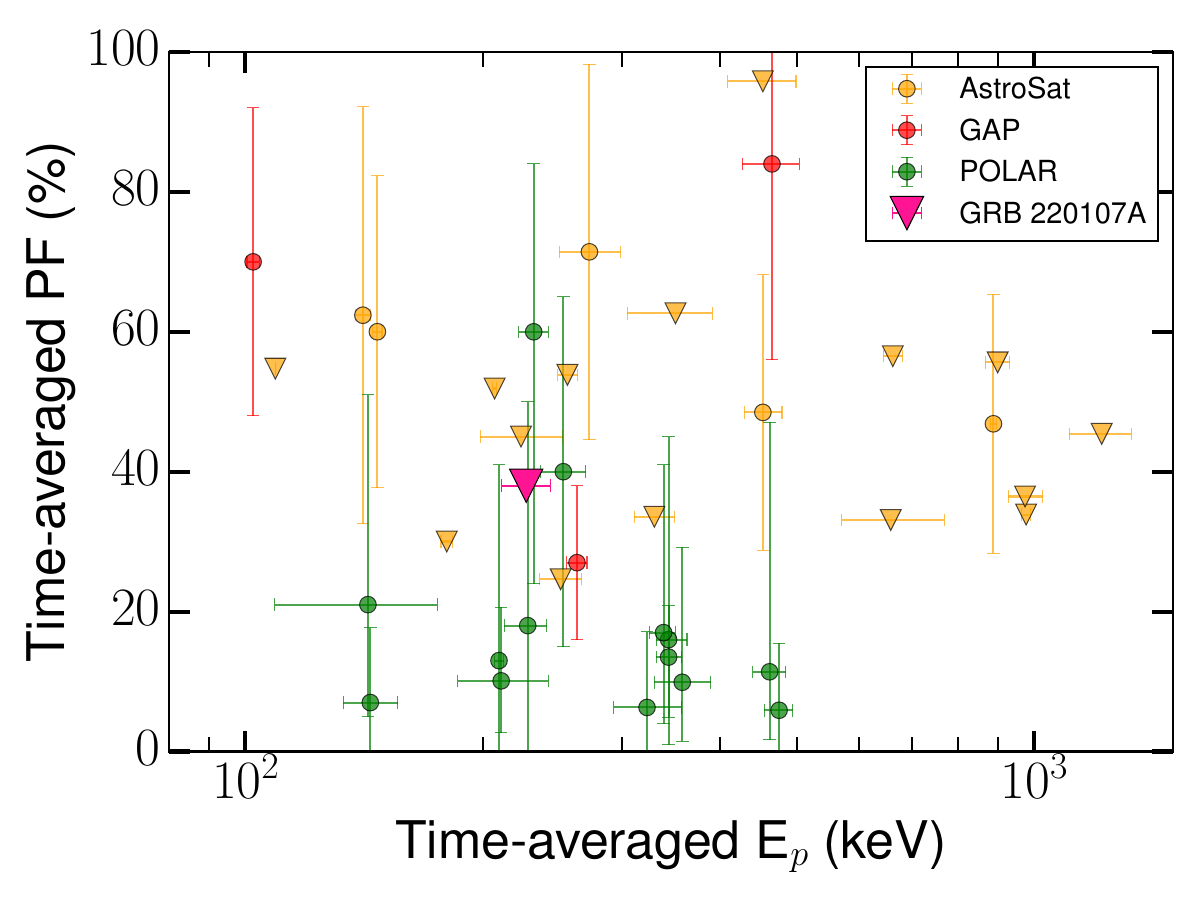}
\includegraphics[width=\hsize]{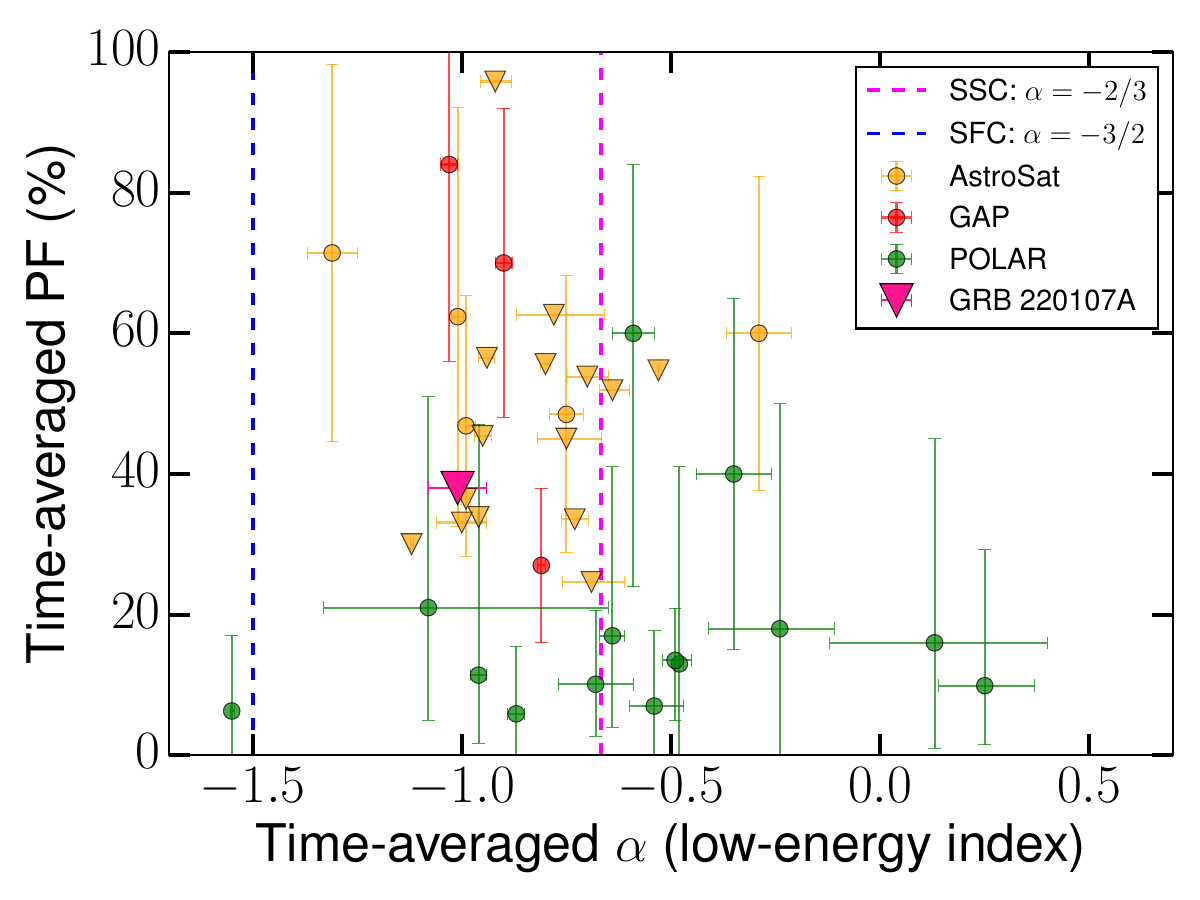}
\caption{Top panel: Distribution of the time-averaged polarization fraction as a function of spectral peak energy for GRB 220107A (pink triangle) compared samples from GAP (red marker), POLAR (green marker), and the \astrosat (orange marker) five-year GRB polarimetry catalog \citep{2011PASJ...63..625Y, 2020A&A...644A.124K, 2022ApJ...936...12C}. No clear trend emerges, consistent with variable prompt-phase polarization in GRBs. Bottom panel: Time-averaged polarization fraction plotted against the low-energy photon index for GRB 220107A (pink triangle) in comparison with GAP, POLAR, and \astrosat measurements. Overlaid dashed lines represent theoretical synchrotron cooling models: pink (synchrotron slow cooling; $\alpha$ = -2/3) and blue (synchrotron fast cooling; $\alpha$ = -3/2).}
\label{fig:TAPCOMPARE}
\end{figure}

To place our results in context, we compared the time-averaged spectro-polarimetric properties of GRB 220107A with those reported by dedicated missions, including GAP, POLAR, and \astrosat/CZTI. In particular, we examined the relation between PF and spectral parameters such as the low-energy photon index ($\alpha$) and the spectral peak energy (\Ep). For this comparison, we used the five-year \astrosat catalog, POLAR catalog, and GAP GRB polarimetry results, and their spectral parameters were obtained from \cite{2011PASJ...63..625Y, 2020A&A...644A.124K, 2022ApJ...936...12C}. We caution that these instruments operate over different energy ranges (e.g., GAP: 50–300 keV, POLAR: 50–500 keV, \astrosat: 100–600 keV) and utilize distinct data analysis frameworks. Consequently, direct comparisons should be treated qualitatively, as instrumental systematics and selection effects may influence the distribution of cataloged parameters. Previous studies have shown that time-averaged PF values in GRBs span a wide range, from undetectable levels ($<10 \%$) to highly polarized cases (up to $\sim$70$ \%$), with no clear correlation between PF and spectral parameters \citep{2022ApJ...936...12C, 2020A&A...644A.124K}. Our results are consistent with this broader observational picture. To visualize these comparisons, we showed two scatter plots incorporating GRB 220107A alongside catalog sources. PF as a function of \Ep on a logarithmic scale and the second plot (see Figure \ref{fig:TAPCOMPARE}) shows PF versus $\alpha$, with theoretical expectations overlaid: dashed pink line for the slow-cooling synchrotron scenario (SSC; $\alpha$ = -2/3) and dashed blue line for the fast-cooling synchrotron scenario (FSC; $\alpha$ = -3/2). For GRB 220107A, the derived \Ep and $\alpha$ place it within the typical Band function parameter space for bright GRBs, yet its $<38 \%$ PF aligns with the majority of catalog entries showing diluted polarization in time-integrated spectra. We noted that the \astrosat/CZTI GRB sample typically exhibits higher PF with a softer low-energy photon index, consistent with prompt emission mechanisms like synchrotron radiation in magnetized outflows. On the other hand, POLAR samples typically show lower PF with a harder low-energy photon index, consistent with mechanisms like photospheric emission in matter-dominated outflows (see Figure \ref{fig:TAPCOMPARE}). While these distinctions highlight potential physical diversity, we cannot rule out that instrumental selection effects contribute to these apparent groupings. This highlights the value of time-resolved spectro-polarimetry, which jointly leverages spectral and polarization information to tightly constrain the physical conditions at the emission site.

\subsubsection{Time-resolved polarization}

To further examine the polarization properties of GRB~220107A, we extended our analysis to include time-resolved spectro-polarimetric measurements. The time-resolved polarimetric study was carried out using \textit{AstroSat}-CZTI data, focusing separately on episodes~1 and~2 of the prompt emission, which had $495$ and $645$ number of Compton events detected, respectively. For episode~1, we obtained a Bayes factor (BF) $<1$, favoring unpolarized radiation, and derived a 1.5-sigma upper limit on the PF of $52\%$, indicating no significant linear polarization and consistent with unpolarized or weakly polarized emission (see Figure \ref{fig:PF} and Table \ref{tab:ERP}). Sliding-window analysis for episode~1 was not performed because the number of Compton events per 12~s interval was below the minimum required ($\sim350$) for a reliable PF estimate. For episode~2, the brighter component also yields BF~$<1$, favoring unpolarized radiation; a two-sigma upper limit of $55\%$ is obtained for the polarization fraction under the assumption of unpolarized emission (see Figure \ref{fig:PF}). To explore variability within episode~2, we employed a sliding-window approach with a 12~s interval shifted in 1~s increments, enabling finer time-resolved PF estimates (see Table \ref{tab:Episode_two_TRP}), shown in pink in the spectro-polarimetric evolution plot (see Figure~\ref{fig:TRS}). The analysis yields nominal best-fit PF values that vary across the episode. However, given the large statistical uncertainties, these variations do not exceed expectations from noise. Consequently, we find no robust evidence for intrinsic temporal evolution in either the polarization fraction or angle (see Table \ref{table:tr_pol}). Nevertheless, since BF~$<3$ in all time bins, no statistically significant detection of polarized emission can be claimed, and only tentative indications can be considered.

\begin{table*}[ht]
\centering
\caption{Polarization measurements of GRB 220107A for the two prompt emission episodes.}
\label{tab:ERP}
\begin{tabular}{ccccccc}
\hline\hline
Time Interval & Energy Band & Compton events & PF (\%) & PA ($^\circ$) & Bayes Factor (BF) \\
(sec) & (keV) &  &  &   \\
\hline
$T_0 - 2$ to $T_0 + 38$ & 100--600 & $\sim 495 $ & $< 52.22 $ (1.5$\sigma$ C.I.) & -- & 0.64  \\
$T_0 + 77$ to $T_0 + 106$ & 100--600 & $\sim 645 $ & $< 54.67$ (2$\sigma$ C.I.) & --  & 0.97\\
\hline
\end{tabular}
\begin{flushleft}
\footnotesize Notes.
Confidence intervals (C.I.) are quoted at the 1.5$\sigma$ and 2$\sigma$ levels. 
\end{flushleft}
\end{table*}

\begin{table*}[ht]
\centering
\caption{Time-resolved polarization measurements of GRB 220107A for the second emission episode. Upper limits are quoted at a 2$\sigma$ confidence level.}
\label{tab:Episode_two_TRP}
\begin{tabular}{cccccc}
\hline\hline
Time Interval & Energy Band & Compton events & PF (\%) & PA ($^\circ$) & Bayes Factor (BF)  \\
(s) & (keV) &  & (1$\sigma$ C.I.) & (1$\sigma$ C.I.) & \\
\hline
$T_0 + 76 $ to $T_0 + 88 $ & 100--600 & 424 & $70 \pm 30$ & $34 \pm 11$ & 2.8 \\
$T_0 + 77$ to $T_0 + 89$   & 100--600 & 502 & $55 \pm26$ & $33 \pm 12$ & 2.3 \\
$T_0 + 78$ to $T_0 + 90$   & 100--600 & 533 & $48 \pm 26$ & $34 \pm 12$ & 1.47 \\
$T_0 + 79$ to $T_0 + 91$   & 100--600 & 549 & $52 \pm 33$ & $33 \pm 12$ & 1.93 \\
$T_0 + 81$ to $T_0 + 93$   & 100--600 & 510 & $47 \pm 30$ & $-26 \pm 16$ & 1.3 \\
$T_0 + 82$ to $T_0 + 94$   & 100--600 & 507 & $53 \pm 30$ & $-26 \pm 14$ & 1.3 \\ \hline
$T_0 + 83$ to $T_0 + 95$   & 100--600 & 507 & < 65  & -- & 0.7 \\
$T_0 + 84$ to $T_0 + 96$   & 100--600 & 510 & < 65 & -- & 0.8 \\
\hline
\label{table:tr_pol}
\end{tabular}
\end{table*}

\subsubsection{Statistical Significance and Detection Thresholds}

The polarization measurements reported in this work must be interpreted within the context of established statistical thresholds for GRB polarimetry. Following standard procedure in the field \citep{2022ApJ...936...12C, 2020A&A...644A.124K}, we assume a Bayes Factor threshold of BF $>$ 3.2 for claiming a positive detection of polarization, corresponding to ``positive evidence" on Jeffreys' scale. Our time-integrated analysis (BF = 0.67) and both episode-level measurements (Episode 1: BF $<$ 1; Episode 2: BF $\sim$ 1) fall well below this threshold, indicating no statistically significant detection of polarization. We therefore report only upper limits for these intervals. 

The highest Bayes Factor obtained in our sliding-window analysis is BF = 2.8 for the interval \fermiT +76 to \fermiT +88 sec, which corresponds to PF = 70 $\pm$ 30\% ($1\sigma$ uncertainty). While this represents a marginally elevated signal compared to other time bins, it still falls below the BF $> $ 3.2 threshold and therefore does not constitute a robust detection by standard criteria. We refer to this as a ``tentative" or ``marginal" signal throughout this work.

The absence of statistically significant polarization detections limits our ability to make definitive claims about emission mechanisms. However, our upper limits and the tentative enhancement in Episode 2 provide valuable constraints: The low polarization in Episode 1 (upper limit $<$ 52\%, at $1.5\sigma$ confidence), viewed in conjunction with the hard low-energy spectral index ($\alpha > -0.6$), is consistent with dissipative photospheric emission scenarios. In these models, the dominant non-thermal component arises from the Comptonization of thermal photons within the optically thick flow; consequently, multiple scatterings suppress polarization across the entire spectral range. The overall low polarization in Episode 2 ($<$ 55\%, at $2\sigma$ confidence) could indicate either unpolarized photospheric emission, synchrotron emission in tangled fields, or ordered-field synchrotron viewed within the jet cone with geometric depolarization.
The marginal elevation to PF = 70 $\pm$ 30\% in the time-resolved bin, while not statistically significant, has a magnitude consistent with theoretical predictions for synchrotron emission in ordered magnetic fields ($\Pi_{\rm syn}$ $\sim$ 70\%; see Section \ref{sec:discussion})

To claim robust detection of evolving polarization mechanisms would require: (i) BF $>$ 3.2 in multiple time bins showing distinct polarization properties; (ii) systematic variation in both PF and PA correlated with spectral changes. Our current dataset does not meet these criteria. Nevertheless, our measurements demonstrate the diagnostic potential of combined spectro-polarimetry beyond simple time-integrated constraints. While previous catalogs (e.g., POLAR, \astrosat) have effectively ruled out high polarization in time-integrated data, such measurements often average over evolving physical conditions. The novelty of GRB 220107A lies in our ability to perform episode-wise analysis on a burst with a clear spectral transition. Thus, rather than just providing another time-integrated upper limit, we provide a direct observational test linking a specific episode-wise spectral feature to its predicted polarimetric signature within a single event. Our results robustly rule out certain extreme scenarios (e.g., very high sustained polarization throughout) and serve as benchmarks for future observations. Thus, GRB 220107A illustrates both the diagnostic promise and current sensitivity limitations of time-resolved spectro-polarimetry.

\begin{figure}[ht]
    \centering
    \includegraphics[width=\hsize]{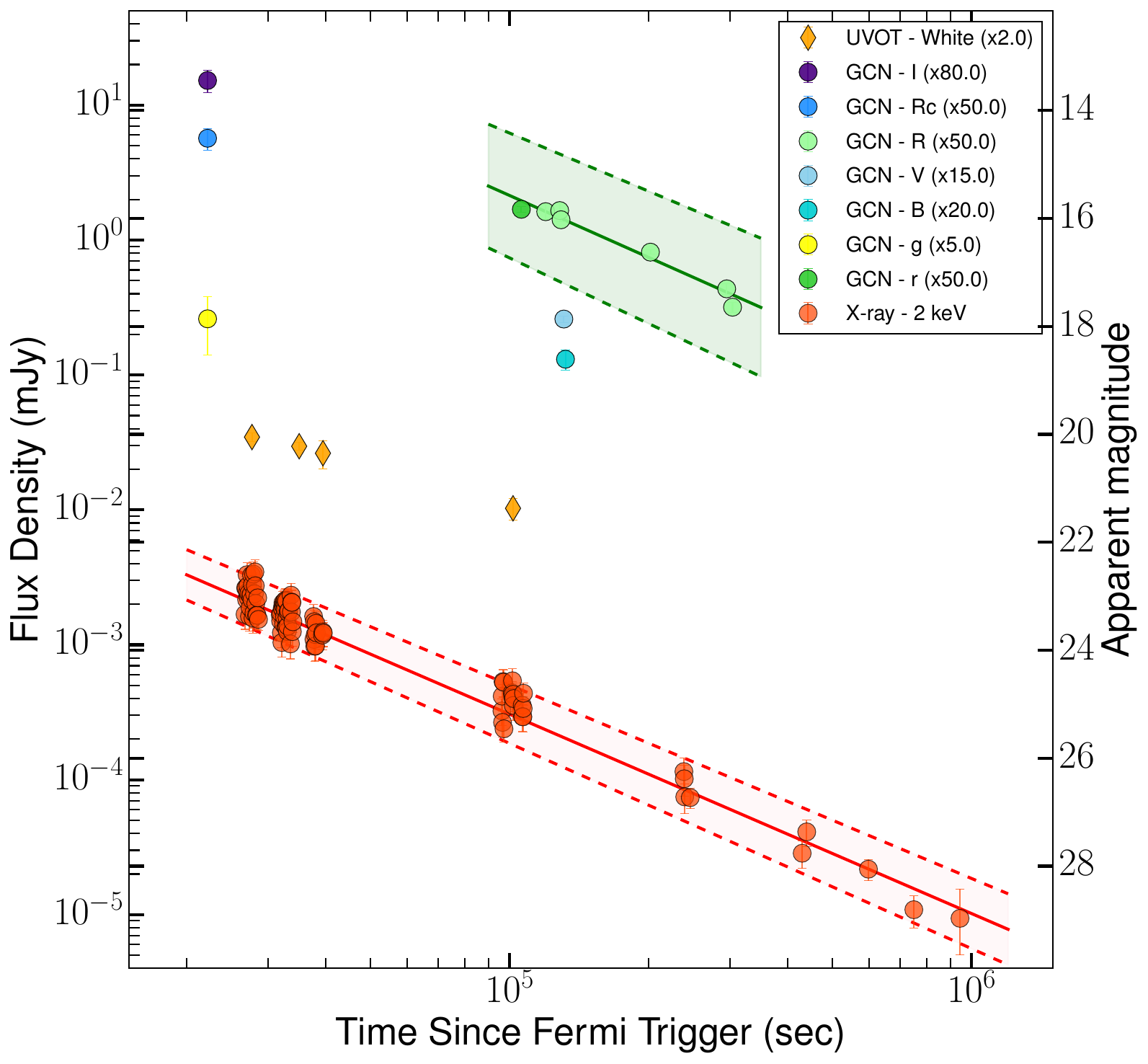}
\caption{Multi-band afterglow light curve of GRB\,220107A. Red points show the \swift/XRT flux density at 2\,keV (right-hand axis, apparent magnitude scale at far right) and orange points show contemporaneous UVOT white-band detections (scaled as indicated in the legend). Colored circles represent ground-based optical photometry from MITSuME Akeno, Nanshan/NEXT, SAO RAS Zeiss-1000, AbAO AS-32, Terskol K-800, and other facilities (individual filter scalings are noted in the legend to improve visibility). Solid lines are single power-law fits to the X-ray (red) and R-band optical (green) data, and the surrounding shaded regions show the 1\,$\sigma$ uncertainty bands from the fit. The X-ray and optical light curves are well described by simple power-law decays with indices $\alpha_{\rm X}=1.48 \pm 0.04$ and $\alpha_{\rm O}=1.52 \pm 0.09$, respectively, indicating an achromatic decline. The figure demonstrates the broadband temporal behavior of the afterglow and the agreement between X-ray and optical decay slopes, which we use to constrain the circumburst environment and emission regime.}
    \label{fig:afterglow}
\end{figure}
\subsection{Afterglow emission of \thisgrb and environment}

To contextualize the prompt emission polarization results, we first characterize the blast-wave energetics and geometry. Specifically, we aim to constrain the jet opening angle ($\theta_{\rm j}$), which is a critical parameter for determining whether the burst was viewed on-axis or off-axis.

The XRT 2\,keV light curve of GRB 220107A is well fit by a single power law with temporal decay index $\alpha_{\rm X}=1.48\pm0.04$, while the R-band optical afterglow decays with $\alpha_{\rm O}=1.52\pm0.09$, i.e., the decay is essentially achromatic within the measurement uncertainties (see Figure \ref{fig:afterglow}). The time-averaged XRT spectrum is described by an absorbed power law with photon index $\Gamma_X=1.87^{+0.09}_{-0.08}$, corresponding to a spectral energy index $\beta_{\rm X}=\Gamma_X-1=0.87^{+0.09}_{-0.08}$. The fitted absorption column, $N_{\rm H}=(2.6\pm1.1)\times10^{21}\,$cm$^{-2}$, exceeds the Galactic foreground ($2.21\times10^{20}\,$cm$^{-2}$; \citealt{2013MNRAS.431..394W}), indicating a modest host/intrinsic contribution along the line of sight. Unless otherwise specified, all uncertainties reported in this section correspond to the $1\sigma$ confidence level.

To identify the blast-wave regime and circumburst medium, we compare the observed temporal and spectral indices with standard fireball closure relations. For a constant-density interstellar medium (ISM) in the slow-cooling regime with $\nu_{\rm m} < \nu < \nu_{\rm c}$, theory predicts \citep{1998ApJ...497L..17S, 2004IJMPA..19.2385Z, 2006RPPh...69.2259M} $\alpha_{\rm X,O}= \tfrac{3}{2}\beta_{\rm X,O}$ (i.e. $\alpha_{\rm X, O}=1.5\beta_{\rm X, O}$), while for frequencies above the cooling break ($\nu > \nu_{\rm c}$) the relation becomes $\alpha_{\rm X,O}=(3\beta_{\rm X,O}-1)/2$ \citep{1999ApJ...519L..17S}. Using the measured $\beta_{\rm X}=0.87^{+0.09}_{-0.08}$, the ISM slow-cooling prediction for $\nu_m<\nu<\nu_c$ is $\alpha_{\rm X, pred}=1.31^{+0.14}_{-0.12}$. This range is fully consistent (within $1\sigma$) with the measured $\alpha_{\rm X}=1.48\pm0.04$ and $\alpha_{\rm O}=1.52\pm0.09$. For the alternative regime $\nu > \nu_{\rm c}$, the expected relation is $\alpha_{\rm X} = (3\beta_{\rm X} - 1)/2 = 0.80^{+0.14}_{-0.12}$, which under-predicts the observed slopes and is therefore disfavored. The wind medium predictions are less consistent: for a wind-like density profile in the slow-cooling $\nu_m<\nu<\nu_c$ regime, one expects  $\alpha_{\rm X,O}=(3\beta_{\rm X,O}+1)/2\approx1.8$--1.9 \citep{2000ApJ...536..195C, 2001ApJ...554..667P}, significantly steeper than observed. The optical decay index ($\alpha_{\rm O}=1.52\pm0.09$) closely matches the X-ray value, supporting an achromatic decay and a common spectral regime for optical and X-rays. Due to the heterogeneous nature of the optical photometric data (collated from GCNs), we do not derive an independent optical spectral index ($\beta_{\rm O}$) and instead rely on the well-constrained X-ray spectral index ($\beta_{\rm X}$) and the achromatic temporal behavior to test the closure relations. The implied electron power-law index is $p=2\beta_{\rm X}+1=2.74^{+0.18}_{-0.16}$. In the ISM slow-cooling regime, this corresponds to $\alpha_{\rm X, O}=3(p-1)/4=1.31^{+0.14}_{-0.12}$, again consistent with the observed values. Thus, the afterglow evolution favors a constant-density circumburst medium with both optical and X-ray frequencies lying below the cooling break.

\section{Discussion}
\label{sec:discussion}

The spectro-polarimetric properties of GRB 220107A provide insights into its radiation mechanisms. In this section, we discuss the results of our spectro-polarimetric analysis.

The theoretical framework for GRB polarization is rooted in the physics of relativistic plasma and magnetic fields. The degree of linear polarization expected from GRBs depends strongly on the emission mechanism and magnetic field geometry. For synchrotron emission, the polarization degree depends on the energy distribution of the emitting electrons ($N(E)$) and the anisotropy of the magnetic field. For synchrotron radiation from electrons with a power-law energy distribution $N(E) \propto E^{-p}$ in a uniform magnetic field, the maximum linear polarization fraction is
\begin{equation}
    \Pi_{\rm syn} = \frac{p+1}{p+\tfrac{7}{3}},
\end{equation}
which yields $\Pi_{\rm syn} \sim 69-75\%$ for typical electron indices $p \approx 2$--3, as commonly inferred from prompt GRB spectral modeling \citep{2002ApJ...581.1248P, 2020NatAs...4..174B, 2015PhR...561....1K}. For a representative value $p \approx 2.5$, this gives $\Pi_{\rm syn} \sim 72\%$ \citep{1979rpa..book.....R}. In realistic scenarios with partially ordered or turbulent magnetic fields, the observed polarization fraction is expected to be lower, typically 10–40\% \citep{2009ApJ...698.1042T}.

For photospheric emission, the polarization depends on the emission process within the hot plasma and the relativistic effects. Models consider mechanisms like Compton scattering in anisotropic radiation fields and the Doppler shift and aberration of emitted photons due to the bulk motion of the plasma \citep{2011ApJ...737...68B, 2014MNRAS.440.3292L, 2014ApJ...783...30C, 2020ApJ...896..139P}. Instead of a uniform signature, the polarization is highly dependent on the viewing angle and the angular structure of the jet (e.g., the Lorentz factor gradient). This sensitivity can manifest as significant temporal variations in the polarization degree and changes in the PA during the burst. 

For photospheric models, the emission originates near the Thomson scattering surface of the jet. In the simplest spherical geometry, the net polarization is zero due to symmetry. Geometric asymmetries or structured jets can lead to modest polarization ($\lesssim 10\%$). However, recent studies \citep{2018ApJ...856..145L, 2020MNRAS.491.3343G} indicate that in `dissipative" photospheres, where sub-photospheric dissipation induces anisotropies, intermediate polarization levels (up to $\sim 40\%$) can be generated. While this allows for measurable polarization, it remains distinct from synchrotron models, which can theoretically accommodate significantly higher maximum values ($> 50\%$) in ordered magnetic fields \citep{2020MNRAS.491.3343G}.

\subsection{Jet Geometry}
\label{sec:geometry}

The observed polarization fraction in GRBs is strongly influenced by the viewing geometry of the jet relative to the observer's line of sight \citep{1999ApJ...524L..43S, 2003ApJ...594L..83G}. To characterize this geometry, we employed the dimensionless parameter $\Gamma \theta_{\rm j}$, which quantifies the product of the bulk Lorentz factor of the outflow ($\Gamma$) and the jet half-opening angle ($\theta_{\rm j}$). This parameter serves as a robust diagnostic of the viewing angle configuration \citep{2005ApJ...618..413G, 2016A&A...594A..84G}: for on-axis observations, where the line of sight lies within the jet cone, $\Gamma \theta_{\rm j} \gg 1$; for off-axis observations, where the line of sight falls outside the jet cone, $\Gamma \theta_{\rm j} \ll 1$ \citep{2020ApJ...896..166R}. A marginal value close to unity corresponds to a narrowly collimated jet seen near its edge \citep{2004IJMPA..19.2385Z}. 

The bulk Lorentz factor can be constrained through multiple independent methods, including temporal and spectral analysis of both prompt and afterglow emission \citep{2010ApJ...725.2209L, 2018A&A...609A.112G}, high-latitude emission delays \citep{2012ApJ...756..112L}, and onset of afterglow \citep{2007A&A...469L..13M}. In this work, we estimate $\Gamma$ for GRB~220107A using the empirical correlation established by \cite{2010ApJ...725.2209L}, which relates the initial bulk Lorentz factor to the isotropic-equivalent $\gamma$-ray energy:\footnote{$\Gamma_{0} \approx 182 \times E_{\gamma, \rm iso, 52}^{0.25 \pm 0.03}$, where $E_{\gamma, \rm iso, 52} = E_{\gamma,\rm iso}/(10^{52}~{\rm erg})$.} This relation has been validated for a large sample of GRBs with well-constrained energetics and has a reported scatter of $\sim 0.3$ dex \citep{2015ApJ...813..116L}. Since the main emission episode was not detected by \fermi/GBM, we obtained $E_{\gamma,\rm iso}$ from \kw observations (see Section~\ref{AMATI}). Applying this relation yields $\Gamma = 392 \pm 36$, consistent with typical values found for long GRBs ($\Gamma \sim 100$--$1000$; \citealt{2006RPPh...69.2259M, 2015PhR...561....1K}).

The jet opening angle was estimated from the \swift/XRT afterglow light curve following the achromatic jet break methodology of \cite{2001ApJ...562L..55F} and \cite{2007MNRAS.380..374P}. It is important to note that this standard relation assumes a homogeneous (``top-hat") jet model with a sharp edge. The jet break time $t_{\rm jet}$ is related to the opening angle through \citep{2001ApJ...562L..55F}:
\begin{equation}
\theta_{\rm j} \approx 0.057 \left(\frac{t_{\rm jet}}{1~{\rm day}}\right)^{3/8} \left(\frac{n_0}{1~{\rm cm}^{-3}}\right)^{1/8} \left(\frac{E_{\rm iso}}{10^{53}~{\rm erg}}\right)^{-1/8} \left(\frac{1+z}{2}\right)^{3/8}~{\rm rad},
\end{equation}
where $n_0$ is the circumburst medium density, $E_{\rm iso}$ is the isotropic kinetic energy, and $z$ is the redshift. Since $\theta_{\rm j}$ depends sensitively on both $n_{0}$ and the electron energy fraction $\epsilon_{e}$, we adopted fiducial values of $n_{0} = 1~{\rm cm}^{-3}$ and $\epsilon_{e} = 0.2$ \citep{2014ApJ...785...29S, 2022JApA...43...11G}, which are characteristic of typical circumburst environments \citep{2011A&A...526A..23S}. These parameters yield $\theta_{\rm j} \gtrsim 0.16$ rad ($\sim 9^\circ$), consistent with the narrow jet opening angles typically inferred for long GRBs ($\theta_{\rm j} \sim 2^\circ$--$20^\circ$; \citealt{2009ApJ...698...43R}).

Combining these estimates, we obtain $\Gamma \theta_{\rm j} \approx 64 \gg 1$ for GRB~220107A. Within the framework of a uniform (top-hat) jet, such a large value strongly suggests that the burst was viewed well within the jet opening angle, i.e., in an on-axis configuration \citep{2005ApJ...630.1003G}. This conclusion is further supported by the detection of bright prompt emission and the early-time afterglow behavior \citep{2005ApJ...630.1003G}. However, realistic GRB jets likely possess angular structure (e.g., Gaussian or power-law profiles) rather than a uniform distribution \citep{2002ApJ...571..876Z, 2004MNRAS.354...86R, 2015MNRAS.450.3549S}. In structured jets, the polarization signature is sensitive to the viewing angle $\theta_{\rm obs}$ relative to the jet core angle $\theta_c$. For a line of sight within the uniform core ($\theta_{\rm obs} < \theta_c$), the polarization properties are similar to the on-axis top-hat case, typically resulting in low polarization due to symmetry. Conversely, viewing angles near the core edge or within the wings of the structure can break this symmetry, potentially enhancing the observed polarization degree \citep{2021Galax...9...82G, 2018MNRAS.478.4128G, 2020MNRAS.491.3343G}. Theoretical models predict that on-axis GRBs should exhibit distinct polarization signatures depending on the dominant radiation mechanism and magnetic field geometry \citep{2020MNRAS.491.3343G}. 

For synchrotron emission from an ordered magnetic field with a specific orientation (e.g., transverse to the shock normal), polarization fractions of $\Pi \sim 70\%$ are expected \citep{1999ApJ...526..697M}, while tangled or randomly oriented fields can suppress the observed polarization to $\Pi \lesssim 10\%$ \citep{2021Galax...9...82G}. Given the high Lorentz factor derived ($\Gamma \sim 392$), the observable region is limited to a narrow cone of width $\sim 1/\Gamma$. While jet structure can induce polarization by breaking the symmetry of the visible patch (e.g., if the line of sight is near the edge of a structured core where gradients are steep; \citealt{2004MNRAS.354...86R}), our deep on-axis geometry ($\Gamma \theta_{\rm j} \gg 1$) implies that the line of sight likely traverses the quasi-homogeneous jet core. In this regime, geometric contributions from jet structure are minimized. Consequently, the observed polarization primarily provides constraints on the magnetic field topology and emission mechanisms responsible for the prompt $\gamma$-ray radiation \citep{2009ApJ...698.1042T}, rather than the global angular profile of the jet.

\subsection{Testing Emission Mechanism Scenarios for \thisgrb}
\label{sec:episodes}

In this section, we use GRB 220107A as a test case to explore how spectro-polarimetric observations can constrain emission mechanisms, while acknowledging the limitations imposed by our marginal polarization measurements. The defining characteristic of GRB 220107A provides the opportunity to test how polarization properties evolve across a drastic spectral transition. The distinct separation of the two episodes allows us to treat them as independent physical regimes. Our approach is to evaluate multiple plausible scenarios against the observational constraints, rather than claiming definitive identification of the dominant mechanism.

{\bf First episode:} 
The first episode shows a relatively hard low-energy photon index ($\alpha > 0.67$), along with a blackbody component. The entire episode is unpolarized, with an upper limit on any intrinsic polarization of $\sim52\%$. These properties are consistent with sub-photospheric dissipation accompanied by quasi-thermal Comptonization, where quasi-thermal radiation advected from deep within the outflow is upscattered by energetic electrons at dissipation sites located at moderate to low optical depths \citep{Iyyani_etal_2015, 2020MNRAS.493.5218S}.
Because the emission undergoes multiple scatterings before escaping the photosphere, any intrinsic polarization is effectively washed out. Moreover, GRB polarization measurements lack spatial resolution; thus, even if locally varying polarization arises from 
scattering, the integrated signal averages out to low values. Consequently, sub-photospheric emission is expected to yield little to no net polarization over the main peak (a few hundred keV) of the spectrum\citep{2018ApJ...856..145L}. The observed spectral cutoff at a few hundred keV can also be explained in this framework, as the Comptonized spectrum tends to thermalize while being advected from the dissipation and subsequently released at the photosphere \citep{Ahlgren_etal_2019}. Overall, the spectro-polarimetric properties of the first episode favor a baryon-dominated outflow, with photospheric emission as the dominant radiation channel.

{\bf Quiescent period and engine activity:} GRB~220107A consists of a $\sim40$~s quiescent interval separating its two main emission episodes. Such a long gap is difficult to explain through internal jet dynamics alone. This suggests a temporary halt in the central engine's energy production after the first episode, before re-igniting to power during the second episode. The renewed activity could signal either a fresh episode of energy injection from the central engine (due to a fallback accretion) or the emergence of emission from a physically distinct region.

{\bf Second episode:} 
The second episode shows a softer low-energy spectral index ($\alpha \sim -0.7$). The polarization analysis of the overall emission yields BF $<$ 1, indicating no significant polarization, though sliding-window analysis hints at possible polarized radiation. In the absence of strong statistical evidence, we consider the following scenarios:
(a) {\it Sub-photospheric dissipation:} A softer $\alpha$ may result if the jet’s kinetic energy is 
dissipated close to the photosphere \citep{2011MNRAS.415.3693R}, if the blackbody component lies outside the fitted energy window, or if the Band $\alpha$ reflects an average over a more complex 
underlying spectral shape \citep{Ahlgren_etal_2019}. In such cases, the net polarization is expected to be negligible when the jet is directed towards the observer.
(b) {\it Optically thin synchrotron in random fields:} A softer $\alpha$ may also reflect synchrotron emission in small-scale, tangled magnetic fields - likely in baryon-dominated jets at the dissipation site above the photosphere. The emission viewed within the jet cone would again yield low to null polarization \citep{2009ApJ...698.1042T,2021Galax...9...82G}.
(c) {\it  Optically thin synchrotron in ordered fields:} If the tentative polarization signal is real, higher polarization is expected when synchrotron emission 
arises in large-scale ordered magnetic fields (coherent at least on the scale of $1/\Gamma$) \citep{2009ApJ...698.1042T,2019ApJ...882L..10S}. This would point to a magnetically dominated jet, suggesting a transition in 
jet composition from radiation-dominated (photospheric emission dominated) in episode 1 to magnetically dominated in episode 2 \citep{2018NatAs...2...69Z}, accompanied by a shift in the dissipation site from below to above the photosphere \citep{2020MNRAS.493.5218S}.

Overall, the hard-to-soft evolution in $\alpha$ is consistent with the general behavior of fireball GRBs \citep{2000ApJS..126...19P, 2006ApJS..166..298K}, where emission is interpreted as initially dominated by photospheric or quasi-thermal components and later evolves toward non-thermal synchrotron radiation. Thus, taken together, the spectra-polarimetric results may suggest an evolution in emission mechanisms between the two episodes, though this interpretation remains tentative given the marginal polarization evidence and inherent model dependencies.

\section{Summary and Conclusion}
\label{sec:conclusion}

We have presented a comprehensive spectro-polarimetric study of GRB 220107A using data from \astrosat/CZTI, \fermi-GBM, and \kw. We use this burst as a test case to demonstrate how time-resolved spectro-polarimetry can constrain prompt emission mechanisms, while also illustrating the current sensitivity limitations of this technique. Our main findings can be summarized as follows:

\begin{itemize}
    \item The time-integrated polarization analysis over the interval \fermiT-2 to \fermiT+106~s shows no statistically significant polarization (PF $< 38\%$, BF = 0.67).  
    \item Unlike time-integrated studies that average over changing physics, our time-resolved analysis suggests distinct behavior across the two emission episodes. During the first episode, the polarization fraction is constrained to a low upper limit ($1.5\sigma$ upper limit $<52\%$), consistent with photospheric emission where multiple scatterings suppress polarization. The second episode also shows a low overall polarization (PF $<55\%$, 2$\sigma$; BF $\sim$ 1), although sliding-window analysis hints at a transient marginally elevated polarization signal (PF = 70 $\pm$ 30 \%) between \fermiT+76 to \fermiT+88 sec, favoring synchrotron emission in the presence of ordered magnetic fields.      
    \item The spectral evolution exhibits a hard-to-soft transition. The low-energy photon index $\alpha$ remains very hard during the first episode, supporting a thermal/photospheric origin, consistent with similar hard spectra seen in the early phases of other bright GRBs. In the second episode, $\alpha$ softens to values around $-0.96 \pm 0.05$ (weighted average), consistent with synchrotron-dominated emission.  

\end{itemize}

Overall, our results suggest that the prompt emission of GRB~220107A likely arises from a combination of photospheric and synchrotron processes, with the dominant mechanism evolving across different emission episodes. Temporal changes in polarization provide important insights into the jet composition and magnetic field structure. This study highlights the critical role of spectro-polarimetric observations in constraining GRB radiation mechanisms.
This burst demonstrates both the promise and current limitations of prompt-phase polarimetry. To definitively test whether multi-episode GRBs show evolving emission mechanisms would require: (i) Higher sensitivity polarimeters achieving BF $>$ 3.2 detections in multiple time bins. (ii) Systematic studies of larger samples to establish whether spectral-polarization correlations are common, and (iii) Simultaneous coverage across broader energy ranges to break spectral degeneracies. Future missions dedicated to X-ray and $\gamma$-ray polarimetry, such as POLAR-2 and COSI, will provide the sensitivity needed (specifically, achieving polarization uncertainties of $\sigma_{\Pi} \lesssim 10\%$ in time-resolved bins) to confirm or refute the tentative trends suggested by cases like GRB 220107A, enabling systematic tests of theoretical models and a deeper understanding of relativistic outflows in multi-episode bursts.

\begin{acknowledgements}
We thank the anonymous referee for their constructive comments and suggestions, which have significantly improved the quality of this manuscript. We thank Prof. A. R. Rao for fruitful suggestions to improve the paper. RG was sponsored by the National Aeronautics and Space Administration (NASA) through a contract with ORAU. The views and conclusions contained in this document are those of the authors and should not be interpreted as representing the official policies, either expressed or implied, of the National Aeronautics and Space Administration (NASA) or the U.S. Government. The U.S. Government is authorized to reproduce and distribute reprints for Government purposes notwithstanding any copyright notation herein. This publication uses data from the AstroSat mission of the Indian Space Research Organisation (ISRO), archived at the Indian Space Science Data Centre (ISSDC). CZT-Imager is built by a consortium of institutes across India, including the Tata Institute of Fundamental Research (TIFR), Mumbai, the Vikram Sarabhai Space Centre, Thiruvananthapuram, ISRO Satellite Centre (ISAC), Bangalore, Inter-University Centre for Astronomy and Astrophysics, Pune, Physical Research Laboratory, Ahmadabad, Space Application Centre, Ahmadabad. The Geant4 simulations for this paper were performed using the HPC resources at IUCAA. This research also has used data obtained through the HEASARC Online Service, provided by the NASA-GSFC, in support of NASA High Energy Astrophysics Programs. S.I. is supported by DST INSPIRE Faculty Scheme (IFA19-PH245). The work of D.F., A.R., D.S. and A.T. is supported by the basic funding program of the Ioffe Institute FFUG-2024-0002. A.T. acknowledges support from the HERMES Pathfinder–Operazioni 2022-25-HH.0 grant.
\end{acknowledgements}

\bibliographystyle{aa} 
\bibliography{GRB220107A} 

\begin{appendix}
\end{appendix}

\end{document}